\documentclass[12pt]{article}
 
\usepackage{amsmath,amssymb}
\usepackage{bbold}
\usepackage{bm}
\usepackage[dvipdfmx]{graphicx}
\usepackage{mediabb}
\usepackage{graphicx, color}
\usepackage{wrapfig}
\usepackage{cite}
\usepackage{here}
\begin{document}
\title{
\begin{flushright}
\ \\*[-80pt] 
\begin{minipage}{0.22\linewidth}
\normalsize
%arXiv:YYMM.NNNN \\
APCTP Pre2020-033
 \\*[50pt]
\end{minipage}
\end{flushright}
{\Large \bf 
Spontaneous  CP violation by modulus $\tau$\\
  in  $A_4$ model of lepton  flavors
%   CP violation caused by modulus $\tau$  \\
%  in  $A_4$  model of lepton flavors
\\*[20pt]}}

\author{ 
\centerline{
Hiroshi Okada $^{a,b}\footnote{E-mail address: hiroshi.okada@apctp.org}$~ and 
~~Morimitsu Tanimoto $^{c}\footnote{E-mail address: tanimoto@muse.sc.niigata-u.ac.jp}$} \\*[5pt]
\centerline{
\begin{minipage}{\linewidth}
\begin{center}
$^a${\it \normalsize
Asia Pacific Center for Theoretical Physics, Pohang 37673, Republic of Korea} \\*[5pt]
$^b${\it \normalsize
Department of Physics, Pohang University of Science and Technology, Pohang 37673,\\ Republic of Korea} \\*[5pt]
$^c${\it \normalsize
Department of Physics, Niigata University, Niigata 950-2181, Japan}
\end{center}
\end{minipage}}
\\*[50pt]}

\date{
\centerline{\small \bf Abstract}
\begin{minipage}{0.9\linewidth}
\medskip 
\medskip 
\small 
We discuss  the modular $A_4$ invariant model of leptons
combining with  the generalized CP symmetry.
In our model,
% CP is conserved for the value of modulus $\tau$ at the imaginary axis, ${\rm Re} \tau=0$ and  the boundary of the fundamental domain.
both CP and modular symmetries are broken spontaneously by the vacuum expectation value of the modulus $\tau$. The source of the CP violation is a non-trivial value 
of ${\rm Re} [\tau]$ while other parameters of the model are real.
The allowed region of $\tau$ is in very narrow one close to the  fixed point $\tau=i$ for both  normal hierarchy (NH) and inverted ones (IH) of neutrino masses. The CP violating Dirac phase $\delta_{CP}$
is predicted clearly in $[98^\circ,\,110^\circ]$ and  $[250^\circ,\,262^\circ]$
for NH at $3\,\sigma$ confidence level.
On the other hand,    $\delta_{CP}$ is in 
$[95^\circ,100^\circ]$ and  $[260^\circ,\,265^\circ]$
for IH at $5\,\sigma$ confidence level.
The predicted $\sum m_i$ is in $[82,\,102]$\,meV for NH
and $\sum m_i=[134,\,180]$\,meV for IH.
The  effective mass 
$\langle m_{ee}\rangle$ for the $0\nu\beta\beta$ decay  is predicted 
in $[12.5,\,20.5]$\,meV and  $[54,\,67]$\,meV for NH and  IH, respectively. 
%%%%%%%%%%%%%%%%%%%%%%%%%%%%%%%%%%%%%%%%%%%%%%%%%%%
\end{minipage}
}

\begin{titlepage}
\maketitle
\thispagestyle{empty}
\end{titlepage}

%%%%%%%%%%%%%%%%%%%%%%%%%%%%%%%%%%%%%%%%%%%%%%%%%%%%%%%%%%%%%%%%%%
%%%%%%%%%%%%%%%%%%%%%%%%%%  Introduction   %%%%%%%%%%%%%%%%%%%%%%% %%%%%%%%%%%%%%%%%%%%%%%%%%%%%%%%%%%%%%%%%%%%%%%%%%%%%%%%%%%%%%%%%%
\section{Introduction}

The non-Abelian discrete symmetries are attractive  ones to understand
flavors of quarks and leptons.
The  $S_3$ flavor symmetry was a pioneer for the quark flavor mixing
 \cite{Pakvasa:1977in, Wilczek:1977uh}.
 It was also discussed  to understand the large mixing angle
 \cite{Fukugita:1998vn} in the oscillation of atmospheric neutrinos \cite{Fukuda:1998mi}. 
 For the last twenty years, the  non-Abelian discrete symmetries of flavors have been developed, that is
  motivated by the precise observation of  flavor mixing angles of  leptons
  \cite{Altarelli:2010gt,Ishimori:2010au,Ishimori:2012zz,Hernandez:2012ra,King:2013eh,King:2014nza,Tanimoto:2015nfa,King:2017guk,Petcov:2017ggy,Feruglio:2019ktm}.
Among them, the $A_4$ flavor model is an attractive one 
because the $A_4$ group is the minimal one including a triplet 
 irreducible representation, 
which allows for a natural explanation of the  
existence of  three families of quarks and leptons 
\cite{Ma:2001dn,Babu:2002dz,Altarelli:2005yp,Altarelli:2005yx,
Shimizu:2011xg,Petcov:2018snn,Kang:2018txu}.
However,  it is difficult to obtain clear predictions of the $A_4$ flavor symmetry because of a lot of free parameters associated with  scalar flavon fields.

Recently, a new  approach to the lepton flavor problem
has been put forward 
based on the invariance under the modular transformation \cite{Feruglio:2017spp}, 
where the model of the finite
modular group  $\Gamma_3 \simeq A_4$ has been presented.
In this approach, fermion matrices are written in terms of modular forms which are   holomorphic functions of  the modulus  $\tau$.
This work  inspired further studies of the modular invariance approach
 to the lepton flavor problem. 
%It should be emphasized that there is a significant difference between the 
%models based on the $A_4$ modular symmetry and those based on the conventional 
% $A_4$ flavor symmetry.
%Yukawa couplings transform non-trivially under the modular group
%and are written in terms of modular forms which are  
%holomorphic functions of a complex parameter,  the modulus  $\tau$.

The finite groups $S_3$, $A_4$, $S_4$, and $A_5$
are realized in modular groups \cite{deAdelhartToorop:2011re}.
Modular invariant flavor models have been also proposed on the $\Gamma_2\simeq S_3$ \cite{Kobayashi:2018vbk},
$\Gamma_4 \simeq S_4$ \cite{Penedo:2018nmg} and  
 $\Gamma_5 \simeq A_5$ \cite{Novichkov:2018nkm}.
%%%%%%%%%%
Phenomenological discussions of the neutrino flavor mixing have been done
based on  $\rm A_4$ \cite{Criado:2018thu,Kobayashi:2018scp,Ding:2019zxk}, $\rm S_4$ \cite{Novichkov:2018ovf,Kobayashi:2019mna,Wang:2019ovr} and 
$\rm A_5$ \cite{Ding:2019xna}.
A clear prediction of the neutrino mixing angles and the CP violating phase was given in  the  simple lepton mass matrices  with
the  $\rm A_4$ modular symmetry \cite{Kobayashi:2018scp}.
On the other hand, the  Double Covering groups  $\rm T'$~\cite{Liu:2019khw,Chen:2020udk}
and $\rm S_4'$ \cite{Novichkov:2020eep,Liu:2020akv} were
realized in the modular symmetry.
Furthermore, modular forms for $\Delta(96)$ and $\Delta(384)$ were constructed \cite{Kobayashi:2018bff},
and the extension of the traditional flavor group  was discussed with modular symmetries \cite{Baur:2019kwi}.
The level $7$ finite modular group $\rm \Gamma_7\simeq PSL(2,Z_7)$
was also presented for the lepton mixing \cite{Ding:2020msi}.
Based on those works, phenomenological studies have been developed  in many works
\cite{deMedeirosVarzielas:2019cyj,
Asaka:2019vev,Asaka:2020tmo,Behera:2020sfe,Mishra:2020gxg,deAnda:2018ecu,Kobayashi:2019rzp,Novichkov:2018yse,Kobayashi:2018wkl,Okada:2018yrn,Okada:2019uoy,Nomura:2019jxj, Okada:2019xqk,
Kariyazono:2019ehj,Nomura:2019yft,Okada:2019lzv,Nomura:2019lnr,Criado:2019tzk,
King:2019vhv,Gui-JunDing:2019wap,deMedeirosVarzielas:2020kji,Zhang:2019ngf,Nomura:2019xsb,Kobayashi:2019gtp,Lu:2019vgm,Wang:2019xbo,King:2020qaj,Abbas:2020qzc,Okada:2020oxh,Okada:2020dmb,Ding:2020yen,Nomura:2020opk,Nomura:2020cog,Okada:2020rjb,Okada:2020ukr,Nagao:2020azf,Nagao:2020snm,Yao:2020zml,Wang:2020lxk,
Abbas:2020vuy}
while theoretical investigations have been also proceeded \cite{Kobayashi:2019xvz,Nilles:2020kgo,Nilles:2020nnc,Kikuchi:2020nxn,
	Kikuchi:2020frp,Ishiguro:2020nuf}.
 %%%%%%%%%%%%%%%%%%%%%%%%%%%%%%%%%%%%%%%%%%%

 In order to test the modular symmetry of flavors, the  prediction
 of the  CP violating Dirac phase is important.
 The CP transformation is non-trivial if the non-Abelian  discrete flavor symmetry is set in the Yukawa sector of a Lagrangian.
 Then, we should discuss  so called the  generalized CP symmetry 
 in the flavor space \cite{Ecker:1981wv,Ecker:1983hz,Ecker:1987qp,Neufeld:1987wa,Grimus:1995zi}.
 It  can  predict the CP violating phase  \cite{Grimus:2003yn}.
 The modular invariance has been also studied  combining with  the generalized CP symmetry in flavor theories \cite{Novichkov:2019sqv,Kobayashi:2019uyt}.
 It provides a powerful framework to predict CP violating phases of quarks and leptons. 
 
In our work, we present the modular $A_4$ invariant model
  with  the generalized CP symmetry.  Both CP and modular symmetries are broken spontaneously by the vacuum expectation value (VEV) of the modulus $\tau$.  
  We discuss the phenomenological implication of this model, that is 
 the Pontecorvo-Maki-Nakagawa-Sakata (PMNS) mixing angles \cite{Maki:1962mu,Pontecorvo:1967fh} 
 and the CP violating Dirac phase  of  leptons,
 which is expected to be observed at T2K and NO$\nu$A experiments \cite{T2K:2020,Adamson:2017gxd}.

%%%%%%%%%%%%%%%%%%%%%%%%%%%%%%%%%%%%%%%%%%%%%%%%%%%%%%%%%%%%%%%%%%%%%%%%%

The paper is organized as follows.
In section 2,  we give a brief review on the generalized CP transformation
 in the modular symmetry. 
In section 3,  we present the CP invariant lepton mass matrix in the $A_4$ modular symmetry.
In section 4, we show the phenomenological implication of our model.
Section 5 is devoted to the summary.
In Appendix A, we present  the tensor product  of the $A_4$ group.
In Appendix B, we show the modular forms  for  weight $2$ and $4$.
In Appendix C,  we show how to determine the coupling coefficients of the charged lepton sector.
In Appendix D , we present how to obtain the Dirac $CP$ phase, the Majorana phases and the effective mass of the $0\nu\beta\beta$ decay.

%%%%%%%%%%%%%%%%%%%%%%%%%%%%%%%%%%%%%%%%%%%%%%%%%%%%%%%%%%%%%%%%%%%%%%
%%%%%%%%%%%%%%%%%%%%%%%%%%%%%%%%%%%%%%%%%%%%%%%%%%%%%%%%%%%%%%%%%%%%%%
\section{Generalized CP transformation in modular symmetry}
\subsection{Generalized CP symmetry}

Let us start with discussing  the generalised $CP$ symmetry 
\cite{Grimus:2003yn,Branco:2011zb}.
The CP transformation is non-trivial if the non-Abelian  discrete flavor symmetry $G$ is set in the Yukawa sector of a Lagrangian.
Let us consider the  chiral superfields.
The CP is a discrete symmetry which involves both Hermitian conjugation of a chiral superfield $\psi(x)$ and inversion of spatial coordinates,
\begin{equation}
\psi(x) \rightarrow {\bf X}_{\bf r}\overline \psi(x_P) \ ,
\label{gCP}
\end{equation}
where $x_P=(t,-{\bf x})$ and ${\bf X_{r}}$ is a unitary transformations
of $\psi(x)$ in the irreducible representation $\bf r$ of the discrete flavor symmetry $G$.   If ${\bf X_{r}}$ is the unit matrix,  the $CP$ transformation is  the trivial one. 
This is the case for the continuous flavor symmetry \cite{Branco:2011zb}.
However, in the framework of the non-Abelian discrete family symmetry,
non-trivial choices of ${\bf X_{r}}$ are  possible.
The unbroken $CP$ transformations of ${\bf X_{r}}$ form the group $H_{CP}$.
Then, ${\bf X_{r}}$ must be consistent with the flavor symmetry transformation,
\begin{equation}
\psi(x) \rightarrow {\rho}_{\bf r}(g)\psi(x) \ , \quad g \in G \ ,
\end{equation}
where ${\rho}_{\bf {r}}(g)$ is the representation matrix for $g$
in the irreducible representation $\bf {r}$.

The consistent condition is obtained as follows.
At first,  perform a $CP$ transformation
$\psi(x) \rightarrow {\bf X}_{\bf r}\overline\psi(x_P)$,
then apply a flavor symmetry transformation, 
$\overline\psi(x_P) \rightarrow {\rho}_{\bf r}^*(g)\overline\psi(x_P)$,
and finally perform an inverse CP transformation.
The whole transformation is written as
$\psi(x) \rightarrow {\bf X}_{\bf r} \rho^*(g) 
{\bf X}^{-1}_{\bf r}\psi(x)$,
which must be equivalent to some flavor symmetry
$\psi(x) \rightarrow {\rho}_{\bf r}(g')\psi(x)$. 
Thus, one obtains  \cite{Holthausen:2012dk}
\begin{equation}
{\bf X}_{\bf r} \rho_{\bf r}^*(g) {\bf X}^{-1}_{\bf r}=
{\rho}_{\bf r}(g') 
\ , \qquad g,\, g' \in G \ .
\label{consistency}
\end{equation}
This equation defines the consistency condition, which has to be respected for consistent implementation of a generalized CP symmetry along with a flavor symmetry \cite{Feruglio:2012cw,Chen:2014tpa}.
This chain $CP\rightarrow g\rightarrow CP^{-1}$ maps the group element $g$ onto
$g'$ and preserves the flavor symmetry group structure. 
That is a homomorphism $v(g)=g'$ of $G$.
Assuming the presence of faithful representations $\bf r$,
Eq.\,(\ref{consistency})  defines a unique mapping of $G$ to itself.
In this case,  $v(g)$ is an automorphism of $G$ \cite{Chen:2014tpa}.

It has been also shown that  
the full symmetry group is isomorphic to a semi-direct product of $G$ and $H_{CP}$,  that is $G\rtimes H_{CP}$, where  $ H_{CP}\simeq \mathbb{Z}_2^{CP}$,
is the group generated by the  generalised $CP$ transformation
under the assumption of $\bf X_{r}$ being a symmetric matrix \cite{Feruglio:2012cw}.

\subsection{Modular symmetry}
The modular group $\bar\Gamma$ is the group of linear fractional transformations
$\gamma$ acting on the modulus  $\tau$, 
belonging to the upper-half complex plane as:
\begin{equation}\label{eq:tau-SL2Z}
\tau \longrightarrow \gamma\tau= \frac{a\tau + b}{c \tau + d}\ ,~~
{\rm where}~~ a,b,c,d \in \mathbb{Z}~~ {\rm and }~~ ad-bc=1, 
~~ {\rm Im} [\tau]>0 ~ ,
\end{equation}
 which is isomorphic to  $PSL(2,\mathbb{Z})=SL(2,\mathbb{Z})/\{\rm I,-I\}$ transformation.
This modular transformation is generated by $S$ and $T$, 
\begin{eqnarray}
S:\tau \longrightarrow -\frac{1}{\tau}\ , \qquad\qquad
T:\tau \longrightarrow \tau + 1\ ,
\label{symmetry}
\end{eqnarray}
which satisfy the following algebraic relations, 
\begin{equation}
S^2 =\mathbb{1}\ , \qquad (ST)^3 =\mathbb{1}\ .
\end{equation}

 We introduce the series of groups $\Gamma(N)$, called principal congruence subgroups, where  $N$ is the level $1,2,3,\dots$.
 These groups are defined by
 \begin{align}
 \begin{aligned}
 \Gamma(N)= \left \{ 
 \begin{pmatrix}
 a & b  \\
 c & d  
 \end{pmatrix} \in SL(2,\mathbb{Z})~ ,
 ~~
 \begin{pmatrix}
  a & b  \\
 c & d  
 \end{pmatrix} =
  \begin{pmatrix}
  1 & 0  \\
  0 & 1  
  \end{pmatrix} ~~({\rm mod} N) \right \}
 \end{aligned} .
 \end{align}
 For $N=2$, we define $\bar\Gamma(2)\equiv \Gamma(2)/\{\rm I,-I\}$.
Since the element $\rm -I$ does not belong to $\Gamma(N)$
% while, since the element $-I$\UTF{0081} does not belong to $\Gamma(N)$,
  for $N>2$, we have $\bar\Gamma(N)= \Gamma(N)$.
   The quotient groups defined as
   $\Gamma_N\equiv \bar \Gamma/\bar \Gamma(N)$
  are  finite modular groups.
In these finite groups $\Gamma_N$, $T^N=\mathbb{1}$  is imposed.
 The  groups $\Gamma_N$ with $N=2,3,4,5$ are isomorphic to
$S_3$, $A_4$, $S_4$ and $A_5$, respectively \cite{deAdelhartToorop:2011re}.

Modular forms $f_i(\tau)$ of weight $k$ are the holomorphic functions of $\tau$ and transform as
\begin{equation}
f_i(\tau) \longrightarrow (c\tau +d)^k \rho(\gamma)_{ij}f_j( \tau)\, ,
\quad \gamma\in G\, ,
\label{modularforms}
\end{equation}
under the modular symmetry, where
  $\rho(\gamma)_{ij}$ is a unitary matrix under $\Gamma_N$.

Superstring theory on the torus $T^2$ or orbifold $T^2/Z_N$ has the modular symmetry \cite{Lauer:1989ax,Lerche:1989cs,Ferrara:1989qb,Cremades:2004wa,Kobayashi:2017dyu,Kobayashi:2018rad}.
Its low energy effective field theory is described in terms of  supergravity theory,
and  string-derived supergravity theory has also the modular symmetry.
Under the modular transformation of Eq.\,(\ref{eq:tau-SL2Z}), chiral superfields $\psi_i$ ($i$ denotes flavors)
transform as \cite{Ferrara:1989bc},
\begin{equation}
\psi_i\longrightarrow (c\tau +d)^{-k_I}\rho(\gamma)_{ij}\psi_j\, .
\label{chiralfields}
\end{equation}

%%%%%%%%%%%%%%%%%%%%%%%%%%%%%%%%%%%%%%%%
 We study global supersymmetric models, e.g., 
minimal supersymmetric extensions of the Standard Model (MSSM).
The superpotential which is built from matter fields and modular forms
is assumed to be modular invariant, i.e., to have 
a vanishing modular weight. For given modular forms 
this can be achieved by assigning appropriate
weights to the matter superfields.
%%%%%%%%%%%%%%%%%%%%%%%%%%
%%%%%%%%%%%%%%%%%%%%%%%%%%%

The kinetic terms  are  derived from a K\"ahler potential.
The K\"ahler potential of chiral matter fields $\psi_i$ with the modular weight $-k$ is given simply  by 
%%%%%%%%%%%%%%%%%%%%%%%%%%%%%%
\begin{equation}
K^{\rm matter} = \frac{1}{[i(\bar\tau - \tau)]^{k}} \sum_i|\psi_i|^2,
\end{equation}
%%%%%%%%%%%%%%%%%%%%%%%%%%%
where the superfield and its scalar component are denoted by the same letter, and  $\bar\tau =\tau^*$ after taking VEV of $\tau$.
%%%%%%%%%%%%%%%%%%%%%%%%%%%%%%%
Therefore, 
the canonical form of the kinetic terms  is obtained by 
changing the normalization of parameters \cite{Kobayashi:2018scp}.
The general K\"ahler potential consistent with the modular symmetry possibly contains additional terms \cite{Chen:2019ewa}. However, we consider only the simplest form of
the K\"ahler potential.
%%%%%%%%%%%%%%%%%%%%%%%%%%%%%%%

For $\Gamma_3\simeq A_4$, the dimension of the linear space 
${\cal M}_k(\Gamma{(3)})$ 
of modular forms of weight $k$ is $k+1$ \cite{Gunning:1962,Schoeneberg:1974,Koblitz:1984}, i.e., there are three linearly 
independent modular forms of the lowest non-trivial weight $2$,
which form a triplet of the $A_4$ group,
 ${\bf Y^{(\rm 2)}_{\bf 3}}(\tau)=(Y_1(\tau),\,Y_2(\tau),\, Y_3(\tau))^T$.
As shown in Appendix A, these modular forms have been explicitly obtained \cite{Feruglio:2017spp}  in the  symmetric base of the 
$A_4$ generators  $S$ and $T$ for the triplet representation:
%%%%%%%%%%%%%%%%%%%%%%%%%%%
\begin{align}
\begin{aligned}
S=\frac{1}{3}
\begin{pmatrix}
-1 & 2 & 2 \\
2 &-1 & 2 \\
2 & 2 &-1
\end{pmatrix},
\end{aligned}
\qquad \qquad
\begin{aligned}
T=
\begin{pmatrix}
1 & 0& 0 \\
0 &\omega& 0 \\
0 & 0 & \omega^2
\end{pmatrix}, 
\end{aligned}
\label{STbase}
\end{align}
%%%%%%%%%%%%%%%%%%%%%%%%%%%%%%%%%
%
where $\omega=\exp (i\frac{2}{3}\pi)$ .

%%%%%%%%%%%%%%%%%%%%%%%%%%%%%%%%%%%%%%%%

\subsection{CP transformation of the modulus $\tau$}
The CP transformation in the modular symmetry  was  given by using the generalized CP symmetry  \cite{Novichkov:2019sqv}.
We summarize the discussion in Ref.\cite{Novichkov:2019sqv} briefly.
Consider the CP and modular transformation $\gamma$ of the chiral superfield $\psi (x)$ assigned to an irreducible unitary representation $\bf r$ of $\Gamma_N$.
The chain $CP\rightarrow \gamma \rightarrow CP^{-1}=\gamma'\in \bar\Gamma$ is expressed as:
  \begin{align}
  \psi(x) &\xrightarrow{\, CP\, } {\bf X}_{\bf r} \overline \psi(x_P)
 \xrightarrow{\ \gamma\ }
 (c\tau^*+d)^{-k} {\bf X}_{\bf r}\, {\rho}_{\bf r}^*(\gamma)\overline\psi(x_P)
 \nonumber\\
 &\xrightarrow{\, CP^{-1}\, }
 (c\tau^*_{CP^{-1}}+d)^{-k} {\bf X}_{\bf r}\, 
 {\rho}_{\bf r}^*(\gamma){\bf X}_{\bf r}^{-1}\psi(x) \,,
\label{chain}
\end{align}
where $\tau_{CP^{-1}}$ is the operation of  $CP^{-1}$  on  $\tau$.
The result of this chain transformation should be equivalent to
a modular transformation $\gamma'$ which maps $\psi(x)$ to $(c'\tau+d')^{-k}  {\rho}_{\bf r}(\gamma')\psi(x)$.
Therefore, one obtains
 \begin{align}
{\bf X}_{\bf r} \rho_{\bf r}^*(\gamma) {\bf X}^{-1}_{\bf r}=
\left (\frac{c'\tau+d'}{c\tau^*_{CP^{-1}}+d}
 \right )^{-k} {\rho}_{\bf r}(\gamma') 
 \, .
\label{consistency2}
\end{align}
Since ${\bf X}_{\bf r}$, $\rho_{\bf r}$ and $\rho_{\bf r'}$ are independent of $\tau$,
the overall coefficient on the right-hand side of Eq.\,(\ref{consistency2})
	has  to be a constant (complex) for non-zero weight  $k$:
	 \begin{align}
	 \frac{c'\tau+d'}{c\tau^*_{CP^{-1}}+d} =\frac{1}{\lambda^*}\, ,
	\label{constant}
	\end{align}
where $|\lambda|=1$ due to the unitarity of 
$\rho_{\bf r}$ and $\rho_{\bf r'}$. 
The values of $\lambda$, $c'$ and $d'$ depend on $\gamma$.

 Taking $\gamma=S$ ($c=1$,\,$d=0$) , and denoting $c'(S)=C$,  $d'(S)=D$
 while keeping  $\lambda(S)=\lambda$, we find
 $\tau=(\lambda\tau^*_{CP^{-1}}-D)/C$ from Eq.\,(\ref{constant}), and consequently,
 \begin{align}
\tau \xrightarrow{\, CP^{-1}\, }
 \tau_{CP^{-1}}=\lambda(C\tau^*+D)\,,  \qquad 
 \tau \xrightarrow{\, CP\, } \tau_{CP}=\frac{1}{C}(\lambda\tau^*-D)\,. 
\label{tauCP1}
\end{align}
Let us act with chain  $CP\rightarrow T \rightarrow CP^{-1}$
on the mudular $\tau$ itself:
 \begin{align}
\tau \xrightarrow{\, CP\, } \tau_{CP}=\frac{1}{C}(\lambda\tau^*-D) 
\xrightarrow{\ T\ } \frac{1}{C}(\lambda(\tau^*+1)-D)
\xrightarrow{\, CP^{-1}\, } \tau+\frac{\lambda}{C} \, .
\label{tauCP2}
\end{align}
The resulting transformation has to be a modular transformation, therefore
$\lambda/C$ is an integer. Since $|\lambda|=1$, we find $|C|=1$ and $\lambda=\pm 1$.
 After choosing the sign of $C$ as $C=\mp 1$ so that ${\rm Im}[\tau_{CP}] >0$,
  the CP transformation of Eq.\,(\ref{tauCP1}) turns to
   \begin{align}
  \tau \xrightarrow{\, CP\, } n-\tau^* \, ,
  \label{tauCP3}
  \end{align}
  where $n$ is an integer.
  The chain  $CP\rightarrow S\rightarrow CP^{-1}=\gamma'(S)$ imposes no furher restrictions on $\tau_{CP}$.
 % Since $S$ and $T$ generate the entire modular group, Eq.\,({\ref{tauCP3}}) is %the most general CP transformation
 % of the modulus $\tau$ compatible with the modular symmetry.
  It is always possible to redefine the CP transformation in such a way that $n=0$
  by using the freedom of  $T$ transformation.
  Therefore, we define that the modulus $\tau$ transforms under CP as
     \begin{align}
   \tau \xrightarrow{\, CP\, } -\tau^* \, ,
   \label{tauCPfinal}
   \end{align}
   without loss of generality.

%%%%%%%%%%%%%%%%%%%%%%%%%
%%%%%  String CP  %%%%%%%
%%%%%%%%%%%%%%%%%%%%%%%%%

The same transformation of $\tau$ was also derived from the higher dimensional theories \cite{Kobayashi:2019uyt}.
%%%%%%%%%%%%%
The four-dimensional CP symmetry can be embedded into $(4+d)$ dimensions as higher dimensional proper Lorentz symmetry with positive determinant.
That is, one can combine the four-dimensional CP transformation and $d$-dimensional transformation with negative determinant so as to obtain $(4+d)$ dimensional proper Lorentz transformation.
For example in six-dimensional theory, we denote the two extra coordinates by a complex coordinate $z$.
The four-dimensional CP symmetry with $z \rightarrow z^*$ or $z \rightarrow -z^*$ is a six-dimensional proper Lorentz symmetry.
Note that $z = x + \tau y$, where $x$ and $y$ are real coordinates.
The latter transformation $z \rightarrow -z^*$ maps the upper half plane ${\rm Im}[\tau]>0$ to the same half plane.
Hence, we consider the transformation $z \rightarrow -z^*$\,$(\tau \rightarrow -\tau^*)$ as the CP symmetry.

%It is remarked that 
% the CP can be  spontaneously broken through the VEV of the modulus %$\tau$. 

%%%%%%%%%%%%%%%%%%%%%%%%%%%%%%%%%%%%%%%%%%%%%%%%%%%%%%%%%%%%%%%%%%%%%
\subsection{CP transformation of  modular multiplets}
Chiral superfields and modular forms   transform  
  in  Eqs.\,(\ref{modularforms}) and (\ref{chiralfields}),
 respectively, under a modular transformation.
 Chiral superfields also  transform
   in  Eq.\,(\ref{gCP})  under the  CP transformation.
  The  CP transformation of modular forms were  given in Ref.\cite{Novichkov:2019sqv} as follows.
  Define a modular multiplet of the irreducible representation $\bf r$
  of $\Gamma_N$   with weight $k$ as $\bf Y^{\rm (k)}_{\bf r}(\tau)$,
  which is transformed as:
  \begin{align}
  \bf Y^{\rm (k)}_{\bf r}(\tau)
  \xrightarrow{\, {\rm CP} \, } Y^{\rm (k)}_{\bf r}(-\tau^*) \, ,
  \end{align} 
under the  CP transformation.
 The complex conjugated CP transformed modular forms
 $\bf Y^{\rm (k)*}_{\bf r}(-\tau^*)$ transform almost like the original multiplets
 $\bf Y^{\rm (k)}_{\bf r}(\tau)$  under a modular transformation, namely:

  \begin{align}
\bf Y^{\rm (k)*}_{\bf r}(-\tau^*) \xrightarrow{\ \gamma \ }
	Y^{\rm (k)*}_{\bf r}(-(\gamma\tau)^*) ={\rm (c\tau+d)^k} 
	\rho_{\bf {r}}^*({\rm u}(\gamma)) 
	Y^{\rm (k)*}_{\bf r}(-\tau^*) \, ,
\end{align}
where $u(\gamma)\equiv CP \gamma CP^{-1}$.
 Using the consistency condition of Eq.\,(\ref{consistency}), we obtain
   \begin{align}
 \bf X_r^T Y^{\rm (k)*}_{\bf r}(-\tau^*) \xrightarrow{\ \gamma \ }
 {\rm (c\tau+d)^k} \rho_{\bf {r}}(\gamma) 
 X_r^T Y^{\rm (k)*}_{\bf r}(-\tau^*) \, .
 \end{align}
 Therefore, if there exist a unique modular multiplet at 
  a level $N$, weight $k$ and representation $\bf r$,
  which is satisfied for $N=2$--$5$ with weight $2$,
  we can express the modular form $\bf Y^{\rm (k)}_{\bf r}(\tau)$ as:
  \begin{align}
\bf Y^{\rm (k)}_{\bf r}(\tau)= {\rm \kappa}  X_r^T Y^{\rm (k)*}_{\bf r}(-\tau^*) \, ,
\label{Yproportion}
\end{align}
where $\kappa$ is a  proportional coefficient.
Since $\bf Y^{\rm (k)}_{\bf r}(-(-\tau^*)^*)=\bf Y^{\rm (k)}_{\bf r}(\tau)$,
Eq.\,(\ref{Yproportion}) gives $\bf X_r^* X_r={\rm |\kappa|^2} \mathbb{1}_r $.
Therefore,
 the matrix $\bf X_r$ is symmetric one, and $\kappa=e^{i \phi}$
 is a phase, which can be absorbed in the normalization of 
 modular forms.
 In conclusion, the CP transformation of modular forms  is given as:
 \begin{align}
\bf Y^{\rm (k)}_{\bf r}(\tau)\xrightarrow{\, {\rm CP} \, }
 Y^{\rm (k)}_{\bf r}(-\tau^*) =X_r  Y^{\rm (k)*}_{\bf r}(\tau)\, .
\end{align} 
It is also emphasized that $\bf X_r=\mathbb{1}_r$ satisfies the consistency
condition Eq.\,(\ref{consistency})
in a basis that  generators of $S$ and $T$ of $\Gamma_N$ are represented by symmetric matrices
because of 
$ \rho^*_{\bf {r}}(S)=  \rho^\dagger_{\bf {r}}(S)= \rho_{\bf {r}}(S^{-1})=
 \rho_{\bf {r}}(S)$ and 
 $ \rho^*_{\bf {r}}(T)=  \rho^\dagger_{\bf {r}}(T)= \rho_{\bf {r}}(T^{-1})$.

The CP transformations of  chiral superfields and modular multiplets
are summalized as follows:
  \begin{align}
 \tau \xrightarrow{\, {\rm CP} \, } -\tau^* \, , \qquad
 \psi (x)  \xrightarrow{\, {\rm CP} \, } X_r \overline \psi (x_P)\, , \qquad
 \bf Y^{\rm (k)}_{\bf r}(\tau)\xrightarrow{\, {\rm CP} \, } 
 Y^{\rm (k)}_{\bf r}(-\tau^*)  =X_r  Y^{\rm (k)*}_{\bf r}(\tau)\, ,
 \label{CPsummary}
 \end{align} 
 where  $\bf X_r=\mathbb{1}_r$ can be taken  in the base of symmetric  generators of $S$ and $T$.
 We use this CP  transformation of modular forms to construct the CP invariant mass matrices in the next section.

%%%%%%%%%%%%%%%%%%%%%%%%%%%%%%%%%%%%%%%%%%%%%%%%%%%%%%%%%%%%%%%%%%%%%
%%%%%%%%%%%%%%%%%%%%%%%%%%%%%%%%%%%%%%%%%%%%%%%%%%%%%%%%%%%%%%%%%%%%%%
%%%%%%%%%%%%%%%%%%%%%%%%%%%%%%%%%%%%%%%%%%%%%%%%%%%%%%%%%%%%%%%%%%%%%%
%%%%%%%%%%%%%%%%%%%%%%%%%%%%%%%%%%%%%%%%%%%%%%%%%%%%%%%%%%%%%%%%%%%%%%
%%%%%%%%%%%%%%%%%%%%%%%%%%%%%%%%%%%%%%%%%%%%%%%%%%%%%%%%%%%%%%%%%%%%%%
\section{CP invariant mass matrix in  $A_4$ modular symmetry}

  Let us discuss the CP invariant lepton mass matrix
  in the framework of the $A_4$ modular symmetry.
  We assign the $A_4$ representation and   weight  for superfields of leptons  
   in Table 1, where
  the three left-handed lepton doublets  compose a $A_4$ triplet $L$,
  and the right-handed charged leptons $e^c$, $\mu^c$ and $\tau^c$ are $A_4$ singlets.
  The weights of the superfields of left-handed leptons and 
   right-handed charged leptons are $-2$ and $0$, respectively.
  Then, the simple lepton mass matrices for  charged leptons and neutrinos
   are obtained \cite{Okada:2020ukr}.
  %%%%%%%%%%%%%%%%%%%%%%%%%%%%%%%%%%%%%%%%%%%%%%%%%%%%% 
  
  \begin{table}[h]
  	\centering
  	\begin{tabular}{|c||c|c|c|c|c|} \hline
  		\rule[14pt]{0pt}{1pt}
  		&$L$&$(e^c,\mu^c,\tau^c)$&$H_u$&$H_d$&$\bf Y_r^{\rm (2)}, 
  		\ \   Y_r^{\rm (4)}$\\  \hline\hline 
  		\rule[14pt]{0pt}{1pt}
  		$SU(2)$&$\bf 2$&$\bf 1$&$\bf 2$&$\bf 2$&$\bf 1$\\
  		\rule[14pt]{0pt}{1pt}
  		$A_4$&$\bf 3$& \bf (1,\ 1$''$,\ 1$'$)&$\bf 1$&$\bf 1$&$\bf 3, \ \{3, 1, 1'\}$\\
  		\rule[14pt]{0pt}{1pt}
  		$k$&$ -2$&$(0,\ 0,\ 0)$ &0&0& \hskip -0.7 cm $2, \qquad 4$ \\ \hline
  	\end{tabular}	
  \caption{ Representations and  weights
  	$k$ for MSSM fields and  modular forms of weight $2$ and $4$.
  	}
  	\label{tb:lepton}
  \end{table}
  %%%%%%%%%%%%%%%%%%%%%%%%%%%%%%%%%%%%%
    The superpotential of the  charged lepton mass term is given in terms of
  modular forms  of weight $2$, $\bf Y^{\rm (2)}_3$.
     It is given as:
     \begin{align}
     w_E&=\alpha_e e^c H_d {\bf Y^{\rm (2)}_3}L+
     \beta_e \mu^c H_d {\bf Y^{\rm (2)}_3}L+
     \gamma_e \tau^c H_d {\bf Y^{\rm (2)}_3}L~,
     \label{chargedlepton}
     \end{align}
     where $L$ is the left-handed $A_4$ triplet leptons.
     We can take real for  $\alpha_e$,  $\beta_e$ and $\gamma_e$.
     Under CP, the superfields transform as:
    \begin{align}
   e^c \xrightarrow{\,  CP\,}\, X_{\bf 1}^* \,\overline  e^c\, , \quad
   \mu^c \xrightarrow{\,  CP\,} X_{\bf 1''}^*\, \overline\mu^c\, , \quad
   \tau^c \xrightarrow{\,  CP\,}\, X_{\bf 1'}^* \,\overline \tau^c\, , \quad
    L \xrightarrow{\,  CP\,}\, X_{\bf 3} \overline  L\, , \quad
    H_d \xrightarrow{\,  CP\,}\,\eta_d\, \overline  H_d\, , 
    \end{align} 
     and we can take $\eta_d=1$ without loss of generality.
     Since the representations of  $S$ and $T$ are symmetric as
     seen in Eq.\,(\ref{STbase}), we can choose $X_{\bf 3}=\mathbb{1}$ 
     and $X_{\bf 1}=X_{\bf 1'}=X_{\bf 1''}=\mathbb{1}$.

     Taking $(e_L, \mu_L,\tau_L)$ in the flavor base,
 % Taking $L=(e_L, \mu_L,\tau_L)$,
    the charged lepton mass matrix $M_E$  is simply written  as:    
%%%%%%%%%%%%%%%%%%%%%%%%%%%%%%%%%%%%%%%%%%%%%%% 
  \begin{align}
   \begin{aligned}
M_E(\tau)=v_d \begin{pmatrix}
\alpha_e & 0 & 0 \\
0 &\beta_e & 0\\
0 & 0 &\gamma_e
\end{pmatrix}
\begin{pmatrix}
Y_1(\tau) & Y_3(\tau) & Y_2(\tau) \\
Y_2(\tau) & Y_1(\tau) & Y_3(\tau) \\
Y_3(\tau) & Y_2(\tau) & Y_1(\tau)
\end{pmatrix}_{RL} \ ,
  \end{aligned}
   \label{ME(2)}
  \end{align}
 %%%%%%%%%%%%%%%%%%%%%%%%%%%%%%%%%%%%%%%%%%%%%%%%
where $v_d$ is  VEV of the neutral component of $H_d$, and coefficients $\alpha_e$, $\beta_e$ and $\gamma_e$ are taken to be  real
without loss of generality.
%This charged lepton mass matrix is the same one in Ref.\cite{Okada:2020ukr}.
Under  CP transformation,  the mass matrix $M_E$ is transformed
following from  Eq.\,(\ref{CPsummary}) as:
\begin{align}
   \begin{aligned}
M_E(\tau)  \xrightarrow{\,  CP\,}  M_E (-\tau^*) = M_E^* (\tau) =
v_d \begin{pmatrix}
		\alpha_e & 0 & 0 \\
		0 &\beta_e & 0\\
		0 & 0 &\gamma_e
	\end{pmatrix}
	\begin{pmatrix}
		Y_1(\tau)^* & Y_3(\tau)^* & Y_2(\tau)^* \\
		Y_2(\tau)^* & Y_1(\tau)^* & Y_3(\tau)^* \\
		Y_3(\tau)^* & Y_2(\tau)^* & Y_1(\tau)^*
	\end{pmatrix}_{RL} \ .
\end{aligned}
\label{CPME}
\end{align}

%%%%%%%%%%%%%%%%%%%%%%%%%%%%%%%%%%%%%%%%%%%%%%%%%%%%%%%%%%%%%
Let us discuss the  neutrino mass matrix.
%%%%%%%%%%%%%%%%%%%%%%%%%%%%%%%%%%%%%
Suppose neutrinos to be Majorana particles.
By using the Weinberg operator, the superpotential of the neutrino mass term, $w_\nu$ is  given as:
\begin{align}
w_\nu&=-\frac{1}{\Lambda}(H_u H_u LL{\bf Y_r^{\rm (4)}})_{\bf 1}~,
\label{Weinberg}
\end{align}
where $\Lambda$ is a relevant cutoff scale.
Since the  left-handed lepton doublet has weight $-2$, the superpotential
is given in terms of  modular forms of weight $4$, ${\bf Y_3^{\rm (4)}}$,
${\bf Y_1^{\rm (4)}}$ and  ${\bf Y_{1'}^{\rm (4)}}$.

By putting $v_u$ for  VEV of the neutral component of $H_u$
%$\langle H_u \rangle =v_u$ and taking $L=(\nu_e, \nu_\mu,\nu_\tau)$ for neutrinos,
and using the tensor products of $A_4$ in Appendix A, we have
\begin{align}
%\begin{aligned}
w_\nu &=\frac{v_u^2}{\Lambda}
\left [ 
\begin{pmatrix}
2\nu_e\nu_e-\nu_\mu\nu_\tau-\nu_\tau\nu_\mu\\
2\nu_\tau\nu_\tau-\nu_e\nu_\mu-\nu_\mu\nu_\tau\\
2\nu_\mu\nu_\mu-\nu_\tau\nu_e-\nu_e\nu_\tau
\end{pmatrix} \otimes
{\bf Y_3^{\rm (4)}}  \right . \nonumber \\
& \left .  + \ 
(\nu_e\nu_e+\nu_\mu\nu_\tau+\nu_\tau\nu_\mu)
\otimes g^{\nu}_1 {\bf Y_1^{\rm (4)}}
+
(\nu_e\nu_\tau+\nu_\mu\nu_\mu+\nu_\tau\nu_e)
\otimes g^{\nu}_2{\bf Y_{1'}^{\rm (4)}}
\right ]  \nonumber \\
=&\frac{v_u^2}{\Lambda}
\left[(2\nu_e\nu_e-\nu_\mu\nu_\tau-\nu_\tau\nu_\mu)Y_1^{(4)}+
(2\nu_\tau\nu_\tau-\nu_e\nu_\mu-\nu_\mu\nu_e)Y_3^{(4)}
+(2\nu_\mu\nu_\mu-\nu_\tau\nu_e-\nu_e\nu_\tau)Y_2^{(4)}\right .
\nonumber \\
& \left .  + \ 
(\nu_e\nu_e+\nu_\mu\nu_\tau+\nu_\tau\nu_\mu)
g^{\nu}_1{\bf Y_1^{\rm (4)}}
+
(\nu_e\nu_\tau+\nu_\mu\nu_\mu+\nu_\tau\nu_e)
g^{\nu}_2{\bf Y_{1'}^{\rm (4)}}
\right ]   \ , 
% \end{aligned}
\end{align}
where ${\bf Y_3^{\rm (4)}}$, ${\bf Y_1^{\rm (4)}}$ and ${\bf Y_{1'}^{\rm (4)}}$
are given in Eq.\,(\ref{weight4}) of Appendix B, and  $g^\nu_{1}$, $g^\nu_{2}$ are complex parameters in general.
The neutrino mass matrix is written as follows:
\begin{align}
M_\nu(\tau)=\frac{v_u^2}{\Lambda} \left [
\begin{pmatrix}
2Y_1^{(4)}(\tau) & -Y_3^{(4)}(\tau) & -Y_2^{(4)}(\tau)\\
-Y_3^{(4)}(\tau) & 2Y_2^{(4)}(\tau) & -Y_1^{(4)}(\tau) \\
-Y_2^{(4)}(\tau) & -Y_1^{(4)}(\tau) & 2Y_3^{(4)}(\tau)
\end{pmatrix}
+g^{\nu}_1 {\bf Y_{1}^{\rm (4)}(\tau)  }
\begin{pmatrix}
1 & 0 &0\\ 0 & 0 & 1 \\ 0 & 1 & 0
\end{pmatrix}
+g^{\nu}_2 {\bf Y_{1'}^{\rm (4)}(\tau) }
\begin{pmatrix}
0 & 0 &1\\ 0 & 1 & 0 \\ 1 & 0 & 0
\end{pmatrix}
\right ] \, ,
\label{neutrinomassmatrix}
\end{align}
which is the same one in Ref.\cite{Okada:2020ukr}.
Under  CP transformation, the mass matrix $M_\nu$ is transformed
following from  Eq.\,(\ref{CPsummary}) as:
\begin{align}
&M_\nu(\tau)  \xrightarrow{\,  CP\,}  M_\nu(-\tau^*) =M^*_\nu(\tau) \nonumber\\
&=\frac{v_u^2}{\Lambda} \left [
\begin{pmatrix}
2Y_1^{(4)*}(\tau) & -Y_3^{(4)*}(\tau) & -Y_2^{(4)*}(\tau)\\
-Y_3^{(4)*}(\tau) & 2Y_2^{(4)*}(\tau) & -Y_1^{(4)*}(\tau) \\
-Y_2^{(4)*}(\tau) & -Y_1^{(4)*}(\tau) & 2Y_3^{(4)*}(\tau)
\end{pmatrix}
+g^{\nu *}_1 {\bf Y_{1}^{\rm (4)*}(\tau)  }
\begin{pmatrix}
1 & 0 &0\\ 0 & 0 & 1 \\ 0 & 1 & 0
\end{pmatrix}
+g^{\nu *}_2 {\bf Y_{1'}^{\rm (4)*}(\tau) }
\begin{pmatrix}
0 & 0 &1\\ 0 & 1 & 0 \\ 1 & 0 & 0
\end{pmatrix}
\right ]  .
\end{align}

In a CP conserving modular invariant theory, both CP and modular symmetries are broken spontaneously by  VEV of the modulus $\tau$.
However, there exist certain values of $\tau$ which conserve CP while breaking the modular symmetry.
Obviously, this is the case if
$\tau$ is left invariant by CP, i.e.
\begin{align}
\tau \xrightarrow{\,  CP\,}   -\tau^*=\tau\, \, ,
\label{CPtau}
\end{align}
which indicates $\tau$ lies on the imaginary axis, ${\rm Re} [\tau]=0$.
In addition to ${\rm Re} [\tau]=0$, 
CP is conserved at the boundary of the fundamental domain.
Then, one has
\begin{align}
 M_E(\tau)=M_E^*(\tau)\, ,\qquad\qquad M_\nu(\tau)=M_\nu^*(\tau) \, ,
\label{CPMassmatrix}
\end{align}
which leads to   $g^{\nu}_1$ and  $g^{\nu }_2$ being  real.
Since parameters 
    $\alpha_e$, $\beta_e$, $\gamma_e$ are also  real,
      the source of the CP violation is only non-trivial ${\rm Re}[\tau] $
    after breaking the modular symmetry.
    In the next section, we present numerical analysis of the CP violation by
    investigating the value of the modulus $\tau$.

%%%%%%%%%%%%%%%%%%%%%%%%%%%%%%%%%%%%%%%%%%%%%%%%%%%%%%%%%%%%
\section{Numerical results of leptonic CP violation}
%%%%%%%%%%%%%%%%%%%%%%%%%%%%%%%%%%%%%%%%%%%%%%%%%%%%%%%%%%%%

 We have presented the CP invariant lepton mass matrices
  in the $A_4$ modular symmetry.
  These mass matrices are the same ones in Ref.\cite{Okada:2020ukr}
  except for  parameters $g_1^\nu$  and $g_2^\nu$ being real.
  If the CP violation will be  confirmed at the experiments of  neutrino oscillations,
  the CP symmetry should be broken spontaneously by  VEV of the modulus $\tau$.
  Thus,  VEV of $\tau$ breaks the CP symmetry as well as the modular invariance.
  % because parameters $g_1^\nu$ and $g_2^\nu$ are real.
  The source of the CP violation is only the real part of $\tau$.
  This situation is different from the previous  work in Ref.\cite{Okada:2020ukr},
  where imaginary parts of  $g_1^\nu$  and $g_2^\nu$  also
  break  the CP symmetry explicitly.
  Our phenomenological concern is whether the spontaneous CP violation
  is realized  due to  the value of $\tau$, which  is consistent with observed lepton mixing angles and neutrino masses.
  If this is the case, the CP violating Dirac phase and Majorana phases
   are predicted clearly under the fixed value of  $\tau$.

   Parameter ratios $\alpha_e/\gamma_e$ and  $\beta_e/\gamma_e$  are given 
   in terms of  charged lepton masses and $\tau$ as shown  in Appendix C.
Therefore, the lepton mixing angles, the Dirac phase  and Majorana phases
are given by our model parameters  $g^{\nu}_1$ and  $g^{\nu }_2$ in addition to the value of $\tau$.

%%%%%%%%%%%%%%%%%%%%%%%%%%%
 As the input charged lepton masses, 
we take Yukawa couplings of charged leptons 
at the GUT scale $2\times 10^{16}$ GeV,  where $\tan\beta=5$ is taken
as a bench mark
\cite{Antusch:2013jca, Bjorkeroth:2015ora}:
\begin{eqnarray}
y_e=(1.97\pm 0.024) \times 10^{-6}, \quad 
y_\mu=(4.16\pm 0.050) \times 10^{-4}, \quad 
y_\tau=(7.07\pm 0.073) \times 10^{-3},
\end{eqnarray}
where lepton masses are  given by $m_\ell=y_\ell v_H$ with $v_H=174$ GeV.

%%%%%%%%%%%%%%%%%%%%%%%%%%%%%%%%%%%%%%%%%%%%%%%%%%%%%%%%%%%%%%%%%%%%%
\begin{table}[H]
	\begin{center}
		\begin{tabular}{|c|c|c|}
			\hline
			      \rule[14pt]{0pt}{2pt}
\ observable\       &                 best fit\,$\pm 1\,\sigma$ for NH                  &           best fit\,$\pm 1\,\sigma$  for IH            \\ \hline
			          \rule[14pt]{0pt}{2pt}				                                          $\sin^2\theta_{12}$               &                     $0.304^{+0.012}_{-0.012}$                     &               $0.304^{+0.013}_{-0.012}$                \\                 \rule[14pt]{0pt}{2pt}
			$\sin^2\theta_{23}$   &                     $0.573^{+0.016}_{-0.020}$                     &               $0.575^{+0.016}_{-0.019}$                \\
			          \rule[14pt]{0pt}{2pt}
			              $\sin^2\theta_{13}$               &                  $0.02219^{+0.00062}_{-0.00063}$            
			              &    $0.02238^{+0.00063}_{-0.00062}$    \\ 
			          \rule[14pt]{0pt}{2pt}
$\Delta m_{\rm sol }^2$  & $7.42^{+0.21}_{-0.20}\times 10^{-5}{\rm eV}^2$  &     $7.42^{+0.21}_{-0.20}\times 10^{-5}{\rm eV}^2$     \\ 
\rule[14pt]{0pt}{2pt}
$\Delta m_{\rm atm}^2$ & \ \ \ \ $2.517^{+0.026}_{-0.028}\times 10^{-3}{\rm eV}^2$ \ \ \ \ & $-2.498^{+0.028}_{-0.028}\times 10^{-3}{\rm eV}^2$ \ \ \\  \hline
		\end{tabular}
\caption{The best fit\,$\pm 1\,\sigma$ of neutrino  parameters from NuFIT 5.0
			for NH and IH 
		\cite{Esteban:2020cvm}.
		}
		\label{DataNufit}
	\end{center}
\end{table}
\vskip -0.5 cm

We also input  the   lepton mixing angles and neutrino mass parameters
which are given by NuFit 5.0 in Table \ref{DataNufit} \cite{Esteban:2020cvm}.
In our analysis, $\delta_{CP}$ is output because its observed range
 is too wide at $3\,\sigma$ confidence level.
We investigate  two possible cases of neutrino masses $m_i$, which are
the normal  hierarchy (NH), $m_3>m_2>m_1$, and the  inverted  hierarchy (IH),
$m_2>m_1>m_3$.
%%%%%%%%%%%%%%%%%%%%%%%%%%%%%%%%%%%%%%%%%%%%%%%%%%%%%%%%%%%%%%%%%%%%%
%%%%%%%%%%%%%%%%%%%%%%%%%%%%%%%%%%%%%%%%%%%%%%%%%%%%%%%%%%%%%%%%%%%%%
Neutrino masses and  
the Pontecorvo-Maki-Nakagawa-Sakata (PMNS) matrix $U_{\rm PMNS}$ \cite{Maki:1962mu,Pontecorvo:1967fh} 
are obtained by diagonalizing 
$M_E^\dagger M_E$ and $M_\nu^\dagger M_\nu$.
We also investigate the effective mass for the $0\nu\beta\beta$ decay,
$\langle m_{ee} \rangle$ (see Appendix D)
and 
the sum of three neutrino  masses  $\sum m_i$  since
it is constrained by the recent cosmological data,
which is  the upper-bound $\sum m_i\leq 120$\,meV obtained at the 95\% confidence level
\cite{Vagnozzi:2017ovm,Aghanim:2018eyx}.
%%%%%%%%%%%%%%%%%%%%%%%%%%%%%%%%%%%%%%%%%%%%%
%%%%%%%%%%%%  Numerical results  %%%%%%%%%%%%
%%%%%%%%%%%%%%%%%%%%%%%%%%%%%%%%%%%%%%%%%%%%%  
\subsection{Case of normal hierarchy of neutrino masses}

  Let us  discuss numerical results for NH of neutrino masses.
  The ratios  $\alpha_e/ \gamma_e$ and $\beta_e/ \gamma_e$  are given 
  after fixing charged lepton masses and $\tau$ as shown  in Appendix C.
 However, in practice, we scan $\alpha_e/ \gamma_e$ and $\beta_e/ \gamma_e$ 
 to obtain the observed  charged lepton mass ratio
  and include them in   $\chi^2$ fit as well as  three mixing angles
   and $\Delta m_{\rm atm}^2/\Delta m_{\rm sol}^2$.

 We have already studied the lepton mass matrices 
 in Eqs.\,(\ref{ME(2)}) and  (\ref{neutrinomassmatrix})
  phenomenologically at the nearby fixed points of the  modulus because
 the spontaneous CP violation in Type IIB string theory  is possibly realized at nearby  fixed points, where the moduli stabilization is performed in a controlled way \cite{Abe:2020vmv,Kobayashi:2020uaj}.
 There are  two  fixed  points in the fundamental domain
 of  $PSL(2,\mathbb{Z})$, $\tau= i$ and  $ \tau =\omega$.
 %The infinite point $\tau = i \infty$ is also a special one
 %in which  the subgroup  $\mathbb{Z}^T_3=\{ I,T,T^2 \}$ of $\rm A_4$ is preserved.
 Indeed, the viable $\tau$  of  our lepton mass matrices
 is found around $\tau=i$ \cite{Okada:2020ukr}.
 
 Based on this result of Ref.\,\cite{Okada:2020ukr}, we scan $\tau$ around $i$ while neutrino couplings
    $g^\nu_{ 1}$  and $g^\nu_{ 2}$ are scanned in the real space of $ [-10,\,10]$.
    As a measure of good-fit, we adopt the sum of one-dimensional 
$\chi^2$ function for four accurately known dimensionless observables
$\Delta m_{\rm atm}^2/\Delta m_{\rm sol}^2$,
$\sin^2\theta_{12}$, $\sin^2\theta_{23}$ and $\sin^2\theta_{13}$
 in NuFit 5.0 \cite{Esteban:2020cvm}. In addition, we employ Gaussian approximations for fitting
 $m_e/m_\tau$ and  $m_\mu/m_\tau$ by using the data of PDG \cite{Zyla:2020zbs}. 
 %%%%%%%%%%%%%%%%%%%%%%%%%%%%%%%%%%  
 %%%%%%%%%%%%%%%%%%%%%%%%%%%%%%%%%% 
 %%%%%%%%%%%%%%%%%%%%%%%%%%%%%%%%%%

  %%%%%%%%%%%%%%%%%%%%%%%%%%%%%%%%%%%%%%%%%%%%%%
  %%%%%%%%%% Numerical Results %%%%%%%%%%%%%%%%%
  %%%%%%%%%%%%%%%%%%%%%%%%%%%%%%%%%%%%%%%%%%%%%%
   In Fig.\,1 we show the allowed region  on the 
  ${\rm Re}\, [\tau]$\,--\,${\rm Im}\,[\tau]$ plane, where three mixing angles and
  $\Delta m_{\rm atm}^2/\Delta m_{\rm sol}^2$ are consistent with observed ones.
  The green, yellow and red regions correspond to $2\sigma$, 
  $3\sigma$ and $5\sigma$ confidence levels, respectively.
  
   %%%%%%%%%%%%%%%%%%%%%%%%%%%%%%%%%%%%%%%%%%%%%
  \begin{figure}[H]
  	\begin{minipage}[]{0.47\linewidth}
  		\vspace{5mm}
  		\includegraphics[{width=\linewidth}]{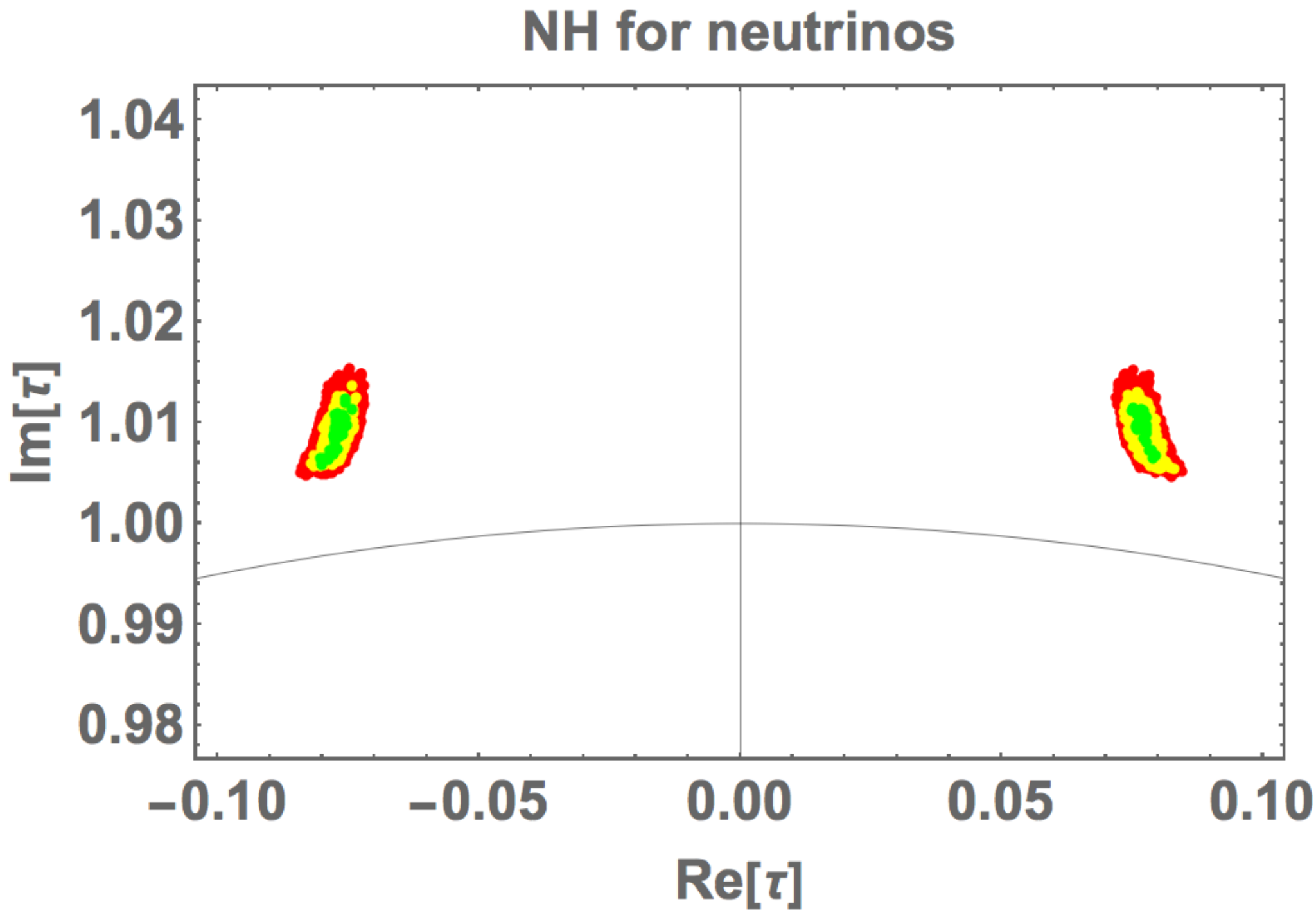}
  		\caption{Allowed regions of $\tau$ for  NH.
  			Green, yellow and red correspond to $2\sigma$, 
  			$3\sigma$, $5\sigma$ confidence levels, respectively.
  			The  solid curve is the boundary of the fundamental domain, $|\tau|=1$.}
  	\end{minipage}
  	\hspace{5mm}
  	\begin{minipage}[]{0.47\linewidth}
  		\vspace{2mm}
  		\includegraphics[{width=\linewidth}]{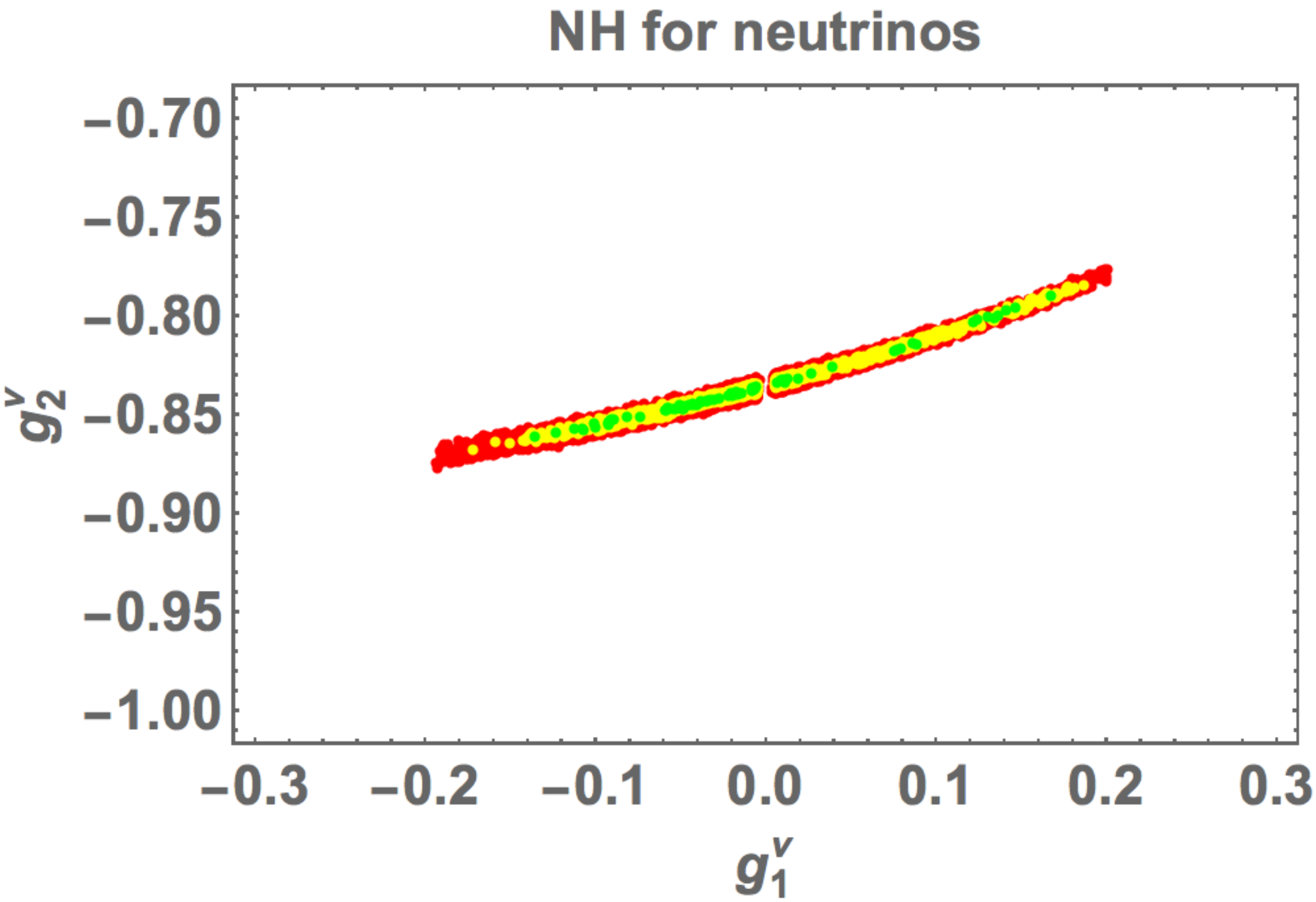}
  		\caption{The allowed region of $g^\nu_1$ and $g^\nu_2$,
  			which are real parameters, for  NH.
  			Colors denote same ones in Fig.\,1.}
  	\end{minipage}
  \end{figure}
  %%%%%%%%%%%%%%%%%%%%%%%%%%%%%%%%%%%%%%%%%%%%%

  The allowed region of $\tau$ is restricted in the narrow regions.
  This result is contrast to the previous one in Ref.\,\cite{Okada:2020ukr},
  where non-trivial phases of $g^\nu_{ 1}$  and $g^\nu_{ 2}$
  enlarged the allowed region of $\tau$.
  The predicted range of  $\tau$ is in ${\rm Re}\,[\tau]=\pm [0.073,0.083]$  and
   ${\rm Im}\,[\tau]=[1.006,1.014]$ at $3\,\sigma$ confidence level (yellow),
   which are close to the fixed point $\tau=i$.

  The allowed region of $g^\nu_1$ and $g^\nu_2$ is also shown in Fig.\,2, where
  $g^\nu_1$ is in the rather wide  region of $[-0.18,0.18]$
  while  $g^\nu_2$ is restricted in  $[-0.87,-0.79]$ at $3\,\sigma$ confidence level (yellow).

 Due to restricted ${\rm Re}\,[\tau]$, the  CP violating Dirac phase $\delta_{CP}$,
 which is defined in Appendix D,  is predicted clearly.
 In Fig.\,3, we show prediction of $\delta_{CP}$ versus the sum of neutrino masses $\sum m_i$.
 It is remarked that  $\delta_{CP}$ is almost independent of $\sum m_i$.
  The predicted ranges of $\delta_{CP}$ are  narrow such as 
  $[98^\circ,110^\circ]$ and  $[250^\circ,262^\circ]$
  at $3\,\sigma$ confidence level (yellow). 
  The predicted ranges $[98^\circ,110^\circ]$ and  $[250^\circ,262^\circ]$
  correspond to ${\rm Re}\,[\tau]= (0.073$--$0.083)$ 
  and ${\rm Re}\,[\tau]=- (0.073$--$0.083)$, respectively. 
  The predicted $\sum m_i$ is in $[82,\,102]$\,meV for $3\,\sigma$ confidence level (yellow).
  The minimal cosmological model, ${\rm \Lambda CDM}+\sum m_i$,
  provides the upper-bound 
  %for the sum of neutrino masses, 
  $\sum m_i<120$\,meV
  \cite{Vagnozzi:2017ovm,Aghanim:2018eyx}. 
  Thus, our predicted  sum of neutrino masses is consistent with 
  the cosmological bound $120$\, meV.
  
% On the other hand, the prediction of the sum of neutrino masses $\sum m_i$
%  depends on $\sin^2\theta_{23}$.

 In Fig.\,4, we show the allowed region on 
  the $\sin^2\theta_{23}$\,--\,$\sum m_i$ plane.
   Since  $\sum m_i$  depends on the value of  $\sin^2\theta_{23}$ significantly,
   the crucial test of our prediction will be available in the near future.
  %%%%%%%%%%%%%%%%%%%%%%%%%%%%%%%%%%%%%%%%%%%%%%%%%%%%%%%%%%%%%%%

 %%%%%%%%%%%%%%%%%%%%%%%%%%%%%%%%%%%%%%%%%%%%%
 \begin{figure}[H]
 	\begin{minipage}[]{0.47\linewidth}
 	%	\vspace{5mm}
 		\includegraphics[{width=\linewidth}]{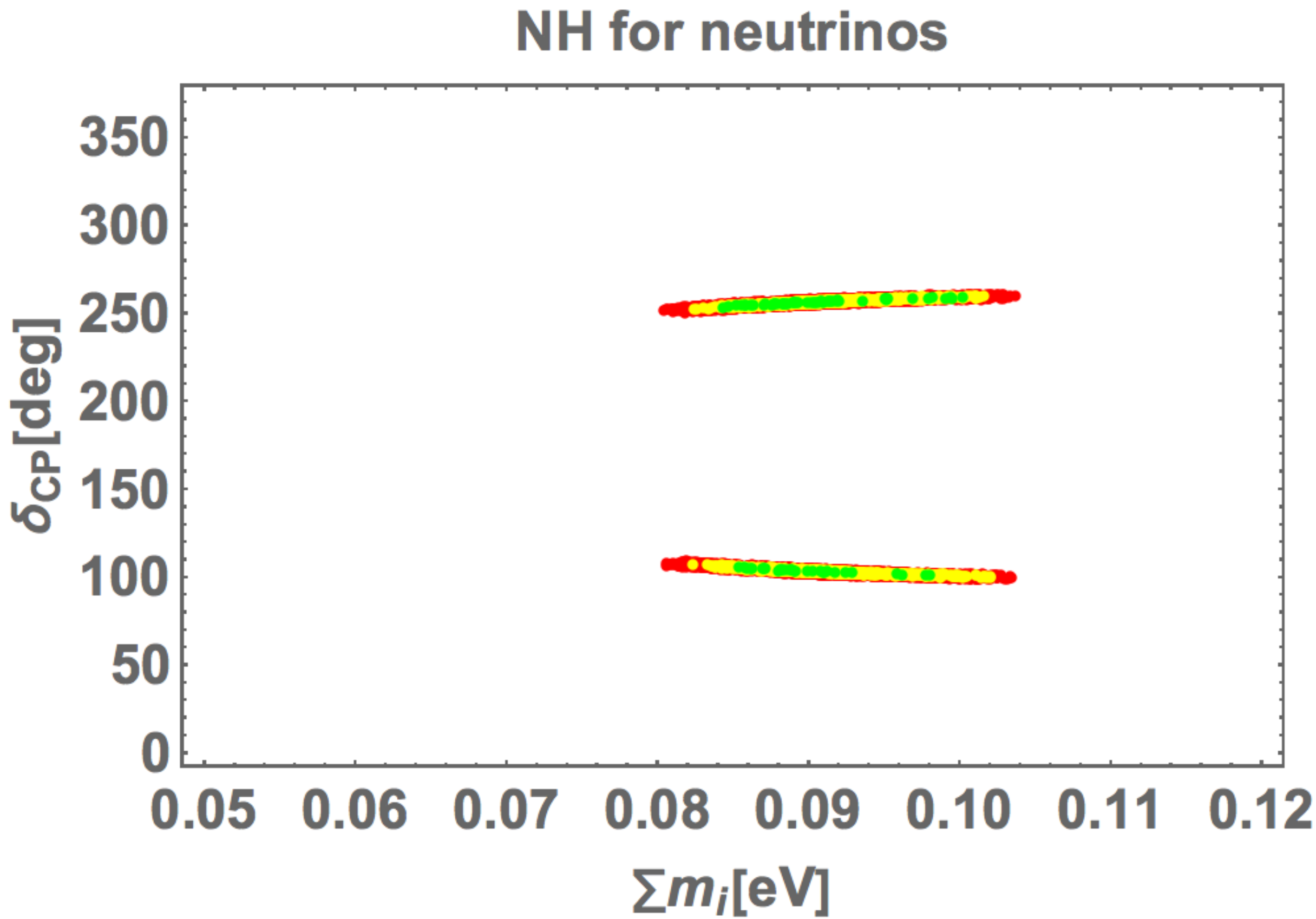}
 \caption{The prediction of $\delta_{CP}$ versus $\sum m_i$
 	for  NH.	Colors denote same ones in Fig.\,1.}
 	\end{minipage}
 	\hspace{5mm}
 	\begin{minipage}[]{0.47\linewidth}
 %		\vspace{-5mm}
 		\includegraphics[{width=\linewidth}]{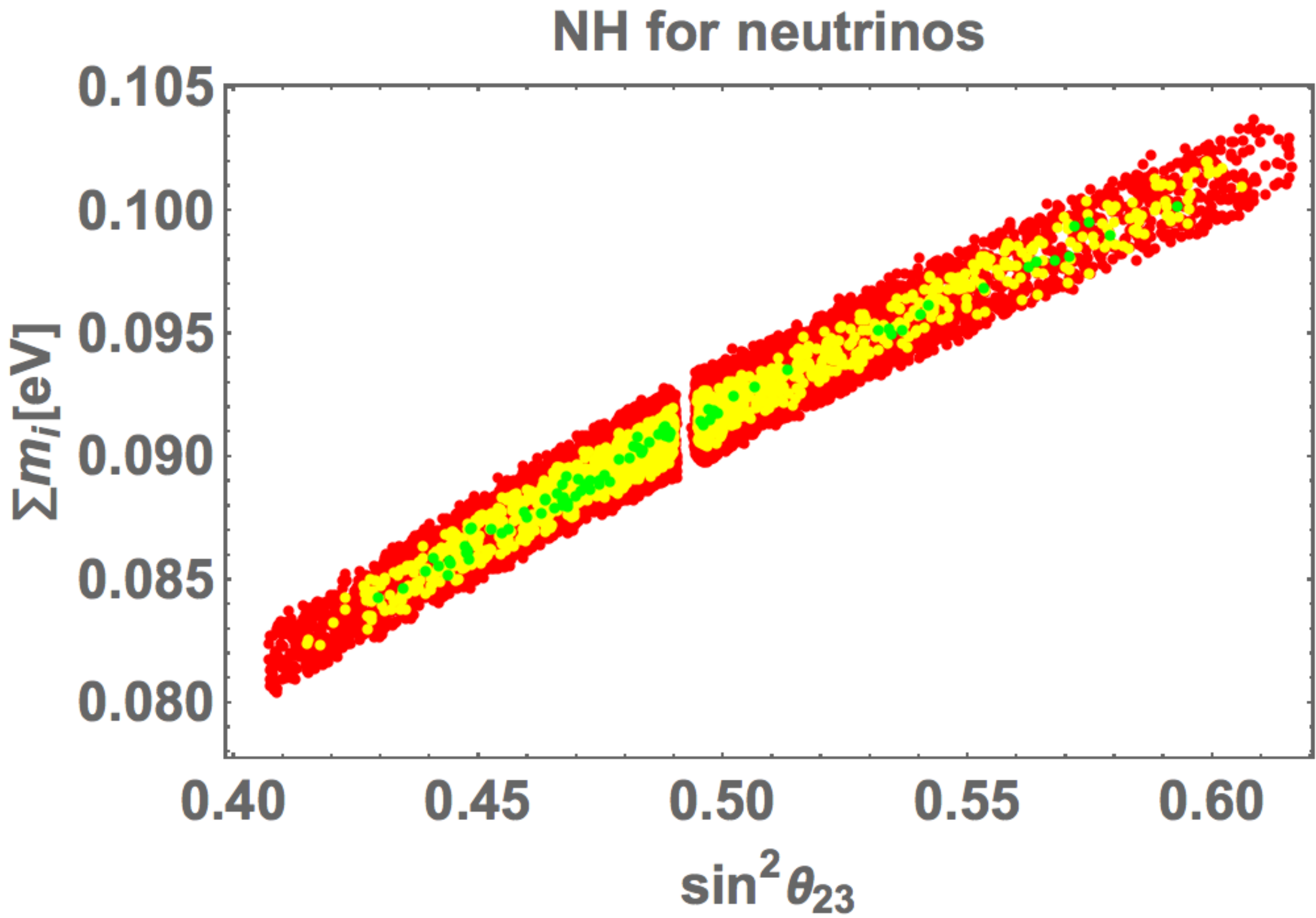}
 \caption{The allowed region on $\sin^2\theta_{23}$--$\sum m_i$
 	 plane for  NH.
 			Colors denote same ones in Fig.\,1.}
 	\end{minipage}
 \end{figure}
 %%%%%%%%%%%%%%%%%%%%%%%%%%%%%%%%%%%%%%%%%%%%%
 
 In Fig.\,5, we show the prediction of Majorana phases $\alpha_{21}$ and $\alpha_{31}$, which are defined by Appendix D. 
 The predicted $[\alpha_{21},\,\alpha_{31}]$ are
 around  $[30^\circ,20^\circ]$
 and $[330^\circ,340^\circ]$ since the source of the CP violation, 
  ${\rm Re}\,[\tau]$ is in the narrow range
    ${\rm Re}\,[\tau]=\pm [0.073,0.083]$.

 We can calculate   the effective mass 
 $\langle m_{ee}\rangle$ for the $0\nu\beta\beta$ decay
  by using  the Dirac phase and Majorana phases as seen in Appendix D.
We show  the predicted value of $\langle m_{ee}\rangle$
versus  $\sin^2\theta_{23}$ as seen in Fig.\,6.
 The predicted $\langle m_{ee}\rangle$ is in $[12.5,\,20.5]$\,meV
 for $3\,\sigma$ confidence level (yellow).
 The prediction of $\langle m_{ee}\rangle\simeq 20$\,meV will be testable in the future experiments  of the neutrinoless double beta decay.

 %%%%%%%%%%%%%%%%%%%%%%%%%%%%%%%%%%%%%%%%%%%%%
 \begin{figure}[H]
 	\begin{minipage}[]{0.47\linewidth}
 	%	\vspace{5mm}
 		\includegraphics[{width=\linewidth}]{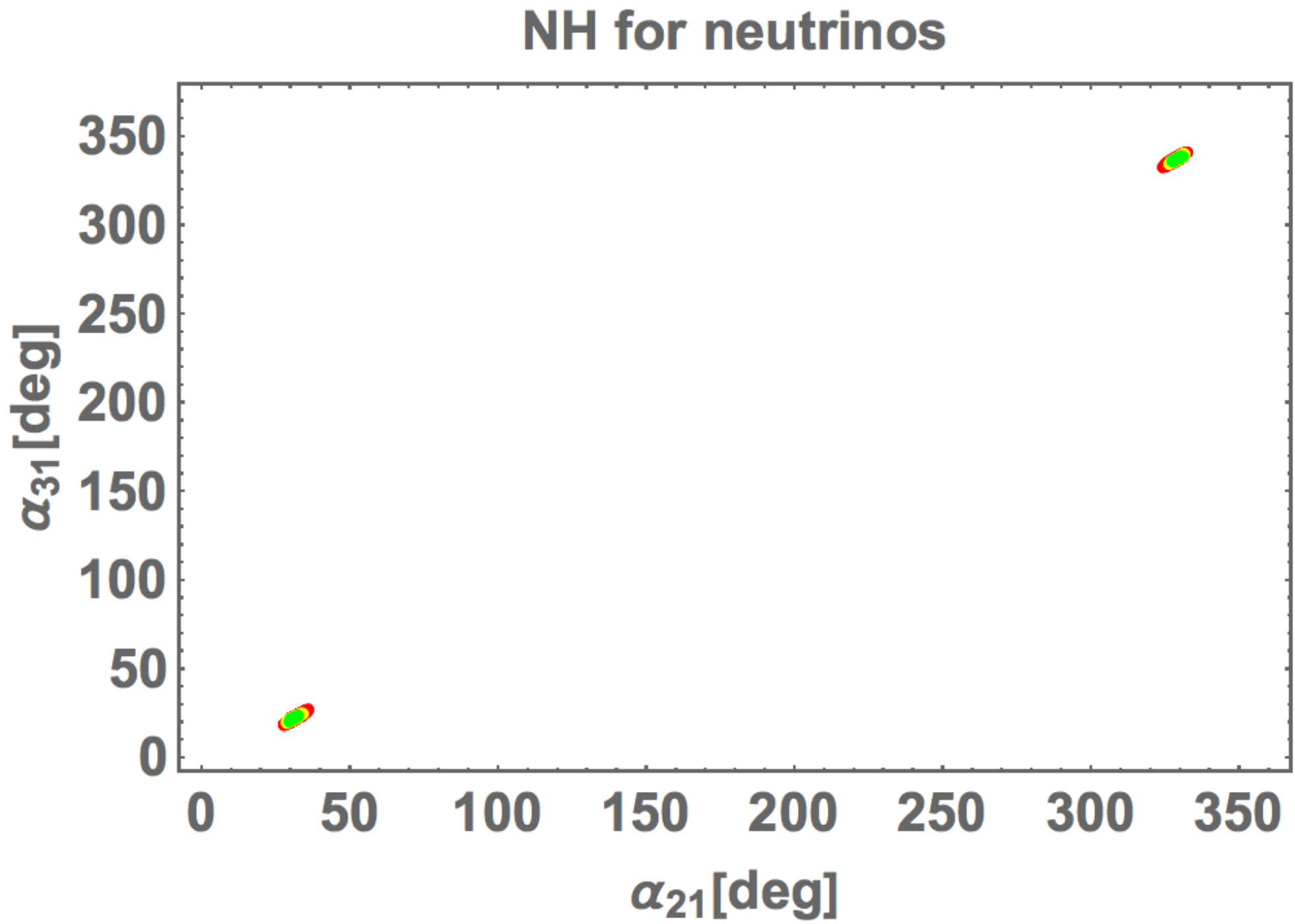}
 		\caption{Predicted Majorana phases $\alpha_{21}$
 			and $\alpha_{31}$ for NH.
 				Colors denote same ones in Fig.\,1.}
 	\end{minipage}
 	\hspace{5mm}
 	\begin{minipage}[]{0.47\linewidth}
 	%	\vspace{-5mm}
 		\includegraphics[{width=\linewidth}]{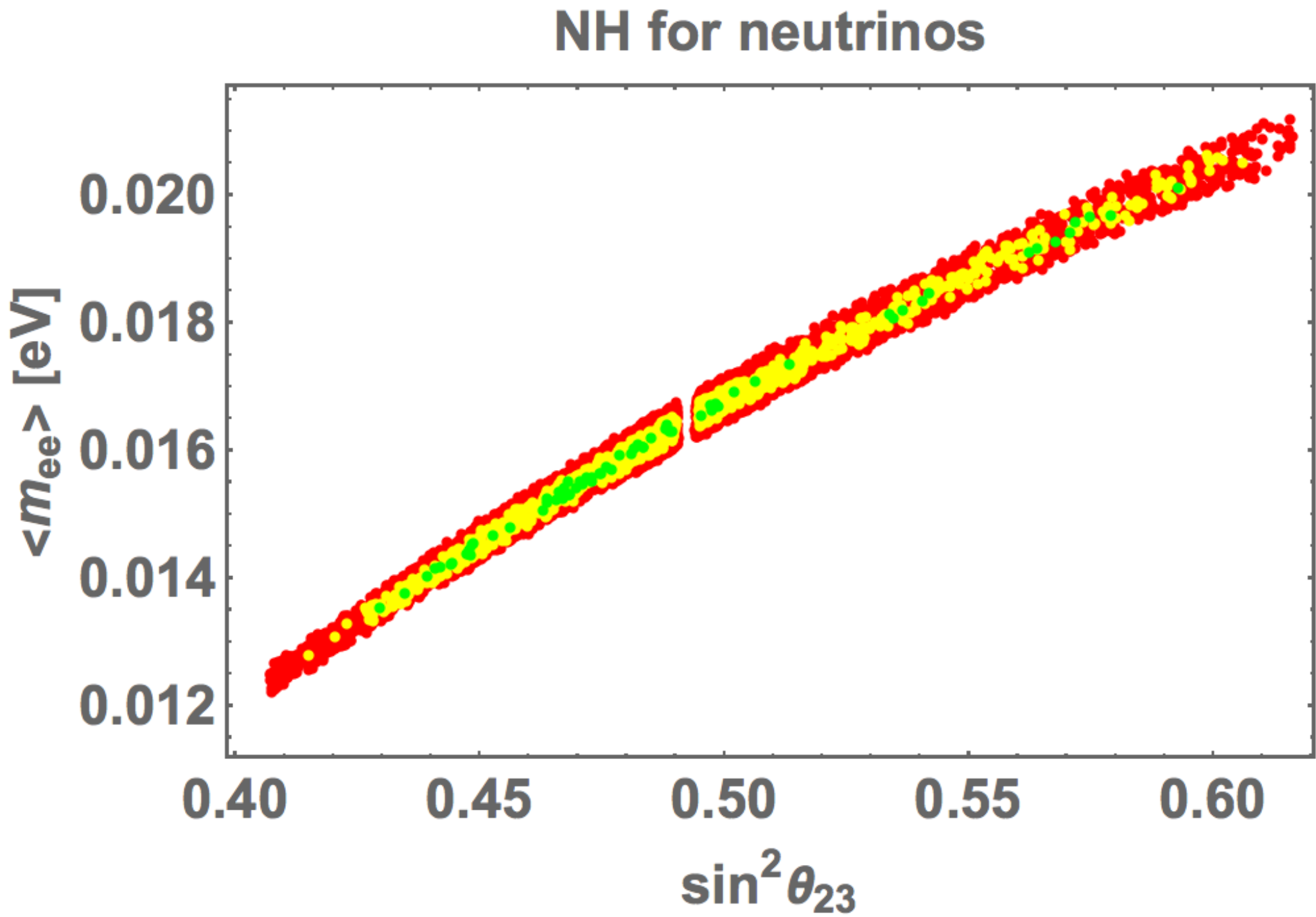}
 		\caption{ The predicted  
 			$\langle m_{ee}\rangle$ versus $\sin^2\theta_{23}$ for  NH.
 				Colors denote same ones in Fig.\,1.}
 	\end{minipage}
 \end{figure}
 %%%%%%%%%%%%%%%%%%%%%%%%%%%%%%%%%%%%%%%%%%%%%

 It is important to understand the difference
 between the results in the present paper and the previous ones
  in  Ref.\cite{Okada:2020ukr},
  where imaginary parts of  $g_1^\nu$  and $g_2^\nu$  also
  break  the CP symmetry explicitly.
  The modulus $\tau$ is severely restricted 
  around  ${\rm Re}\,[\tau]=\pm 0.08$  and
  ${\rm Im}\,[\tau]=1.01$ in this work while 
  it is allowed in rather wide region in the previous work.
  Indeed,  the samller ${\rm Re}\,[\tau]$ and 
   the larger   ${\rm Im}\,[\tau]$ are  allowed
   such as ${\rm Re}\,[\tau]\simeq \pm 0.03$ and 
    ${\rm Im}\,[\tau]\simeq 1.1$ in the previous results.
  Due to this restricted $\tau$ in this work,
  $\delta_{CP}$ and the sum of neutrino masses $\sum m_i$
  are predicted clearly. On the other hand, 
  the CP conservation is still allowed and $\sum m_i$ could be
  larger than $120$\,meV  in the previous work. 
   Moreover, the Dirac phase $\delta_{CP}$ depends on $\sum m_i$.

%%%%%%%%%%%%%%%%%%%%%%%%%%%%%%%%%%%%%%%%%%%%%%%%%%%%%%%%%%
%%%%%%%%%%%%%%%%%%%%%%%%%%%%%%%%%%%%%%%%%%%%%%%%%%%%%%%%%%
%%%%%%%%%%%%%%%%%%%%%%%%  IH  %%%%%%%%%%%%%%%%%%%%%%%%%%%%
%%%%%%%%%%%%%%%%%%%%%%%%%%%%%%%%%%%%%%%%%%%%%%%%%%%%%%%%%%
%%%%%%%%%%%%%%%%%%%%%%%%%%%%%%%%%%%%%%%%%%%%%%%%%%%%%%%%%%
\subsection{Case of inverted  hierarchy of neutrino masses}

We discuss  the case of IH of neutrino masses.
In Fig.\,7, we show the allowed region  on the 
${\rm Re}\, [\tau]$\,--\,${\rm Im}\, [\tau]$ plane, where
the  red region corresponds to $5\,\sigma$ confidence level like in Fig.\,1.
However, there are no  green and yellow regions of $2\,\sigma$ and $3\,\sigma$  confidence levels.

%%%%%%%%%%%%%%%%%%%%%%%%%%%%%%%%%%%%%%%%%%%%%
\begin{figure}[H]
	\begin{minipage}[]{0.47\linewidth}
		\vspace{5mm}
		\includegraphics[{width=\linewidth}]{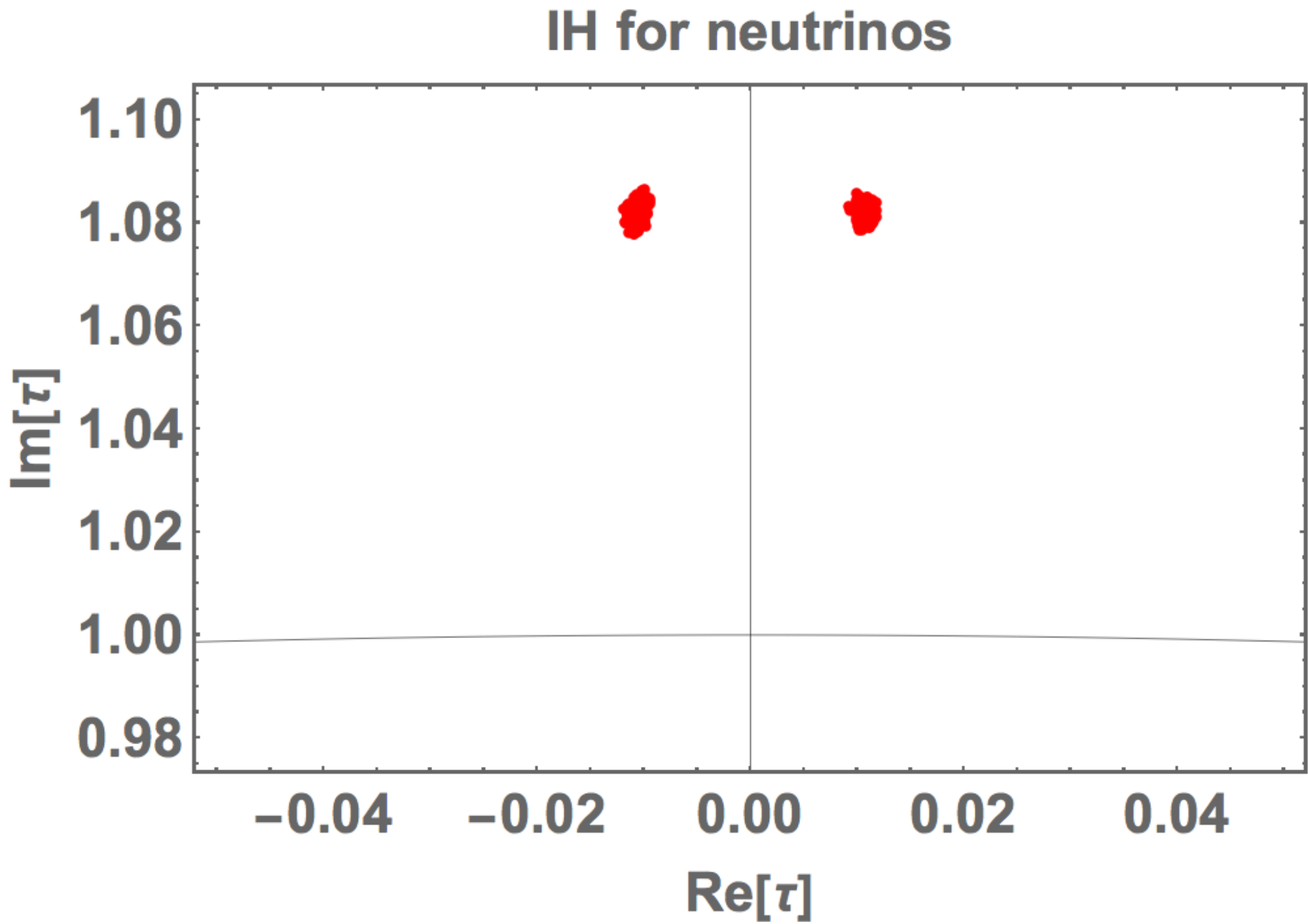}
		\caption{Allowed regions of $\tau$ for  IH.
			Red corresponds  to $5\,\sigma$ confidence level.
% The solid curve is the boundary of the fundamental domain, $|\tau|=1$.
}
	\end{minipage}
	\hspace{5mm}
	\begin{minipage}[]{0.47\linewidth}
		\vspace{2mm}
		\includegraphics[{width=\linewidth}]{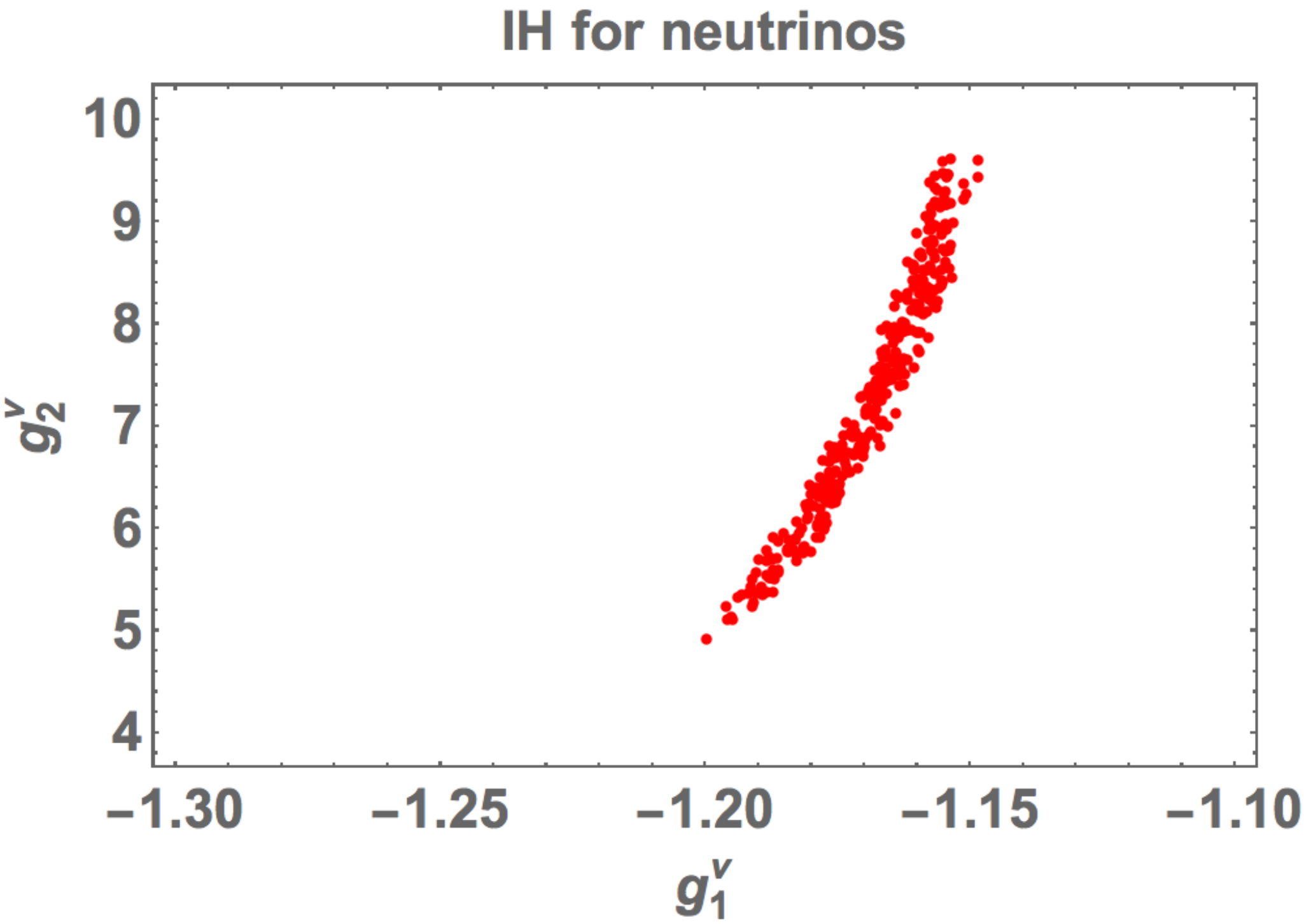}
		\caption{The allowed region of $g^\nu_1$ and $g^\nu_2$,
			which are real parameters, for  IH.}
	\end{minipage}
\end{figure}
%%%%%%%%%%%%%%%%%%%%%%%%%%%%%%%%%%%%%%%%%%%%%
%%%%%%%%%%%%%%%%%%%%%%%%%%%%%%%%%%%%%%%%%%%%%
%%%%%%%%%%%%%%%%%%%%%%%%%%%%%%%%%%%%%%%%%%%%%

The range of  $\tau$ is in ${\rm Re}\,[\tau]=\pm [0.009,\,0.012]$  and
${\rm Im}\,[\tau]=[1.076,\,1.087]$ at $5\,\sigma$ confidence level,
which are close to  $\tau=i$.

The allowed region of $g^\nu_1$ and $g^\nu_2$ is also  shown in Fig.\,8, where
$g^\nu_1$ is restricted in the narrow range of  $[-1.20,\,-1.15]$
while  $g^\nu_2$ is rather large as in  $[4.8,\,9.6]$ for $5\,\sigma$.

In Fig.\,9, we show prediction of $\delta_{CP}$ versus $\sum m_i$.
It is remarked that  $\delta_{CP}$ is almost independent of $\sum m_i$.
The predicted range of $\delta_{CP}$ is in 
$[95^\circ,100^\circ]$ and  $[260^\circ,265^\circ]$
at $5\,\sigma$ confidence level 
while the sum of neutrino masses are in the range of $[134,\,180]$\,meV.
In our numerical result, there is  no region of the sum of neutrino masses
less than $120$\,meV.
The upper-bound  of the minimal cosmological model,
${\rm \Lambda CDM}+\sum m_i$,
is $\sum m_i<120$\,meV \cite{Vagnozzi:2017ovm,Aghanim:2018eyx}, however, 
it becomes  weaker when the data are analysed in the context of extended cosmological models \cite{Zyla:2020zbs}.
The  predicted  sum of neutrino masses of IH
may be  still consistent with the cosmological bound.
%%%%%%%%%%%%%%%%%%%%%%%%%%%%%%%%%%%%%%%%%%%%%
\begin{figure}[h]
	\begin{minipage}[]{0.47\linewidth}
		%	\vspace{5mm}
		\includegraphics[{width=\linewidth}]{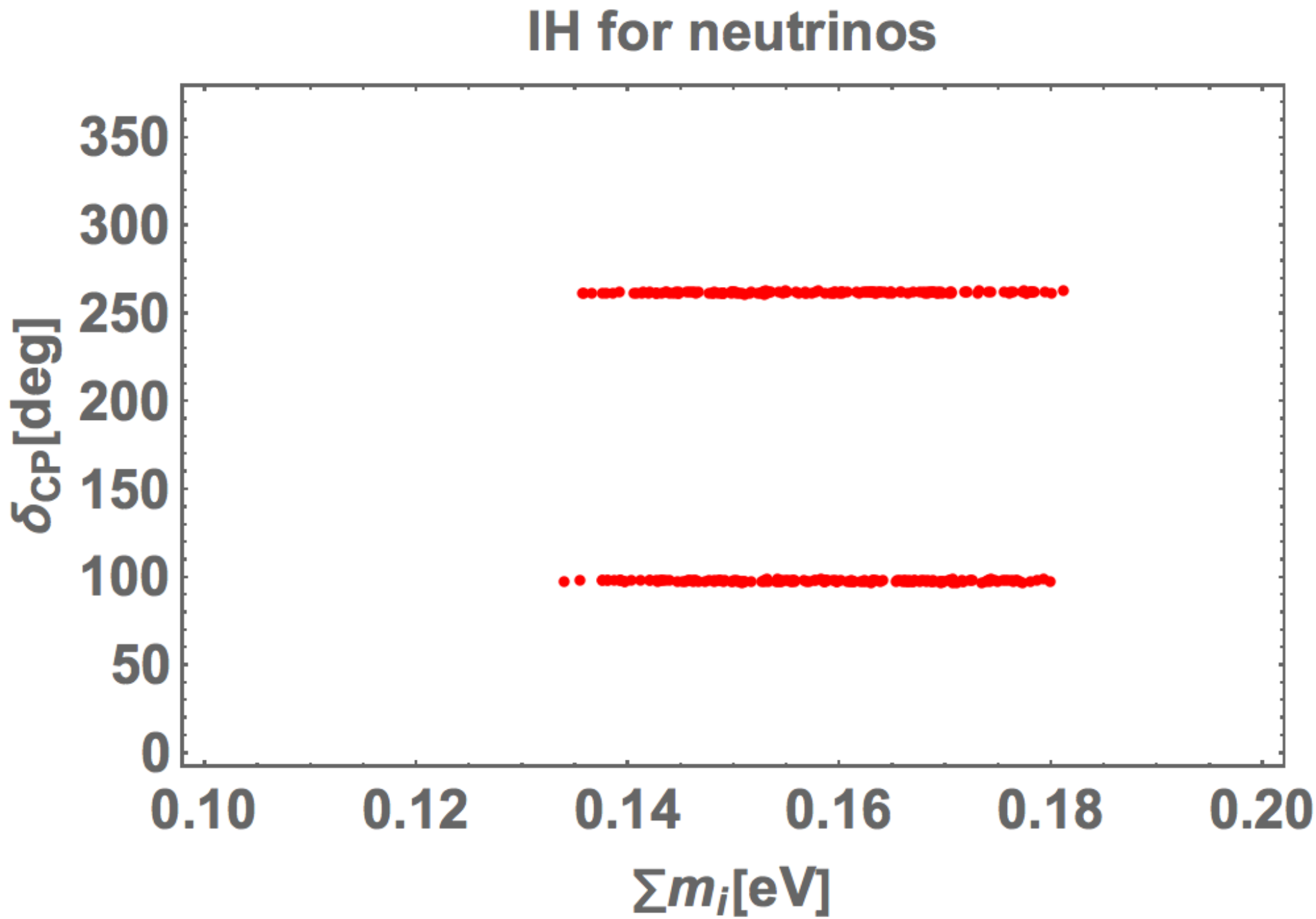}
		\caption{The prediction of $\delta_{CP}$ versus  $\sum m_i$
			for  IH.}
	\end{minipage}
	\hspace{5mm}
	\begin{minipage}[]{0.47\linewidth}
		%		\vspace{-5mm}
		\includegraphics[{width=\linewidth}]{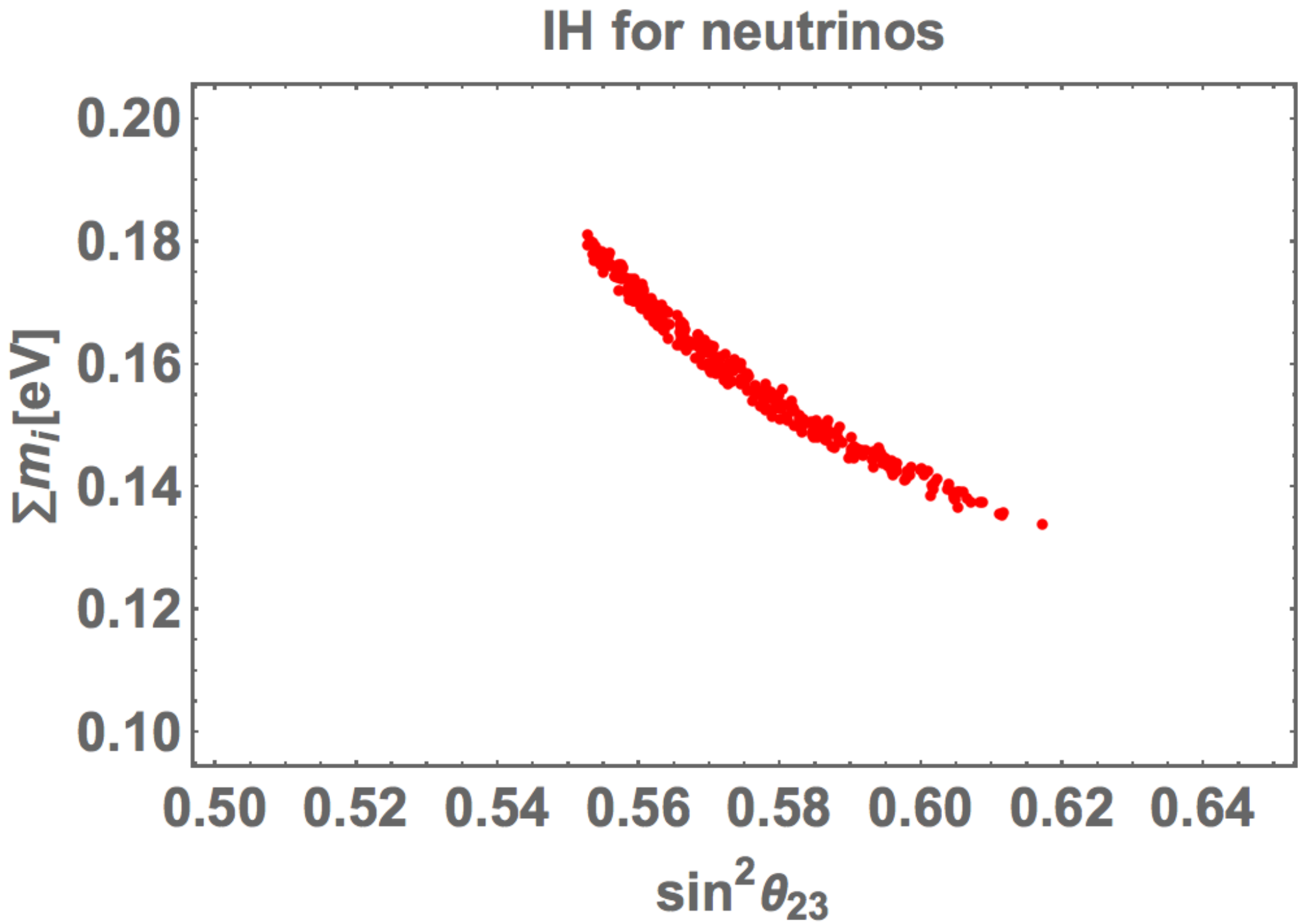}
		\caption{The allowed region on $\sin^2\theta_{23}$--$\sum m_i$
			plane for IH.}
	\end{minipage}
\end{figure}
%%%%%%%%%%%%%%%%%%%%%%%%%%%%%%%%%%%%%%%%%%%%%

We show the allowed region on the $\sum m_i$\,--\,$\sin^2\theta_{23}$ plane in Fig.\,10. The precise measurement of $\sin^2\theta_{23}$
 will provide a severe test for our prediction
 since $\sin^2\theta_{23}>0.55$ is obtained for IH.

In Fig.\,11, we show the prediction of Majorana phases $\alpha_{21}$ and $\alpha_{31}$. 
The predicted $[\alpha_{21},\,\alpha_{31}]$ are
restricted around $[3^\circ,182^\circ]$
and $[356^\circ,178^\circ]$.
We also show  the predicted value of $\langle m_{ee}\rangle$
versus  $\sin^2\theta_{23}$ as seen in Fig.\,12.
The predicted $\langle m_{ee}\rangle$ is in  $[54,\,67]$\,meV
for $5\,\sigma$ confidence level.

%%%%%%%%%%%%%%%%%%%%%%%%%%%%%%%%%%%%%%%%%
%%%%%%%%%%%%%%%%%%%%%%%%%%%%%%%%%%%%%%%%%
As well as the case of NH, we comment on the difference
between the results in the present paper and the previous ones
in  Ref.\cite{Okada:2020ukr},
where $g_1^\nu$ and $g_2^\nu$ are complex.
 Our results are obtained  at more than  $3\,\sigma$ confidence level,
 on the other hand, the previous ones are  at less than $3\,\sigma$ confidence level.
The modulus $\tau$ is also severely restricted in this work while 
it is allowed in rather wide region in the previous work.
The sum of neutrino masses $\sum m_i$ is  lager than $120$\,meV
in this work, on the other hand, it is allowed to be smaller than
$120$\,meV in the previous work. For example, it could be $90$\,meV, and 
the Dirac phase $\delta_{CP}$ depends on $\sum m_i$.
%%%%%%%%%%%%%%%%%%%%%%%%%%%%%%%%%%%%%%%%

%%%%%%%%%%%%%%%%%%%%%%%%%%%%%%%%%%%%%%%%%%%%%
\begin{figure}[H]
	\begin{minipage}[]{0.47\linewidth}
		%	\vspace{5mm}
		\includegraphics[{width=\linewidth}]{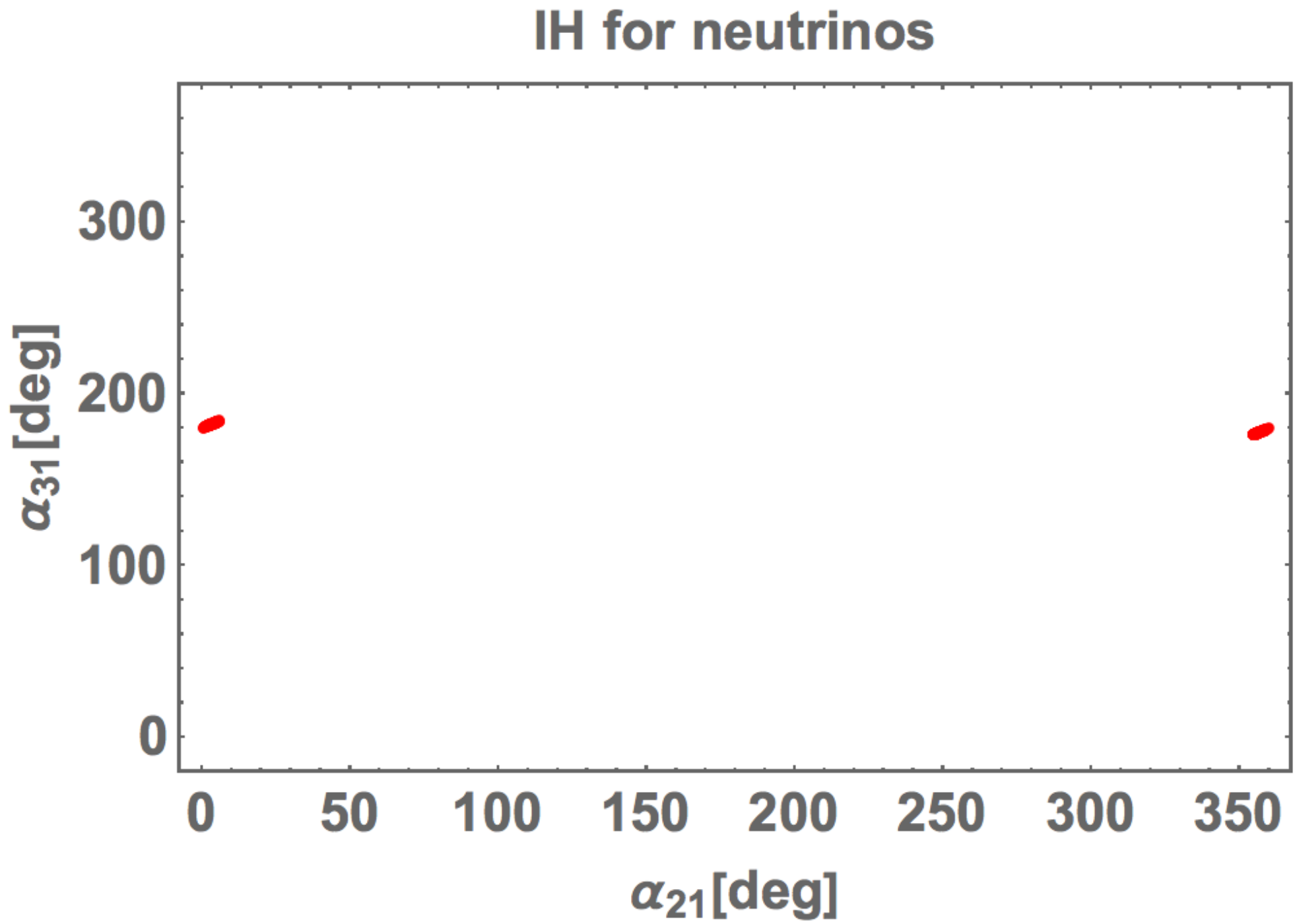}
		\caption{Predicted Majorana phases  $\alpha_{21}$
			and $\alpha_{31}$ for IH.}
	\end{minipage}
	\hspace{5mm}
	\begin{minipage}[]{0.49\linewidth}
		%	\vspace{-5mm}
		\includegraphics[{width=\linewidth}]{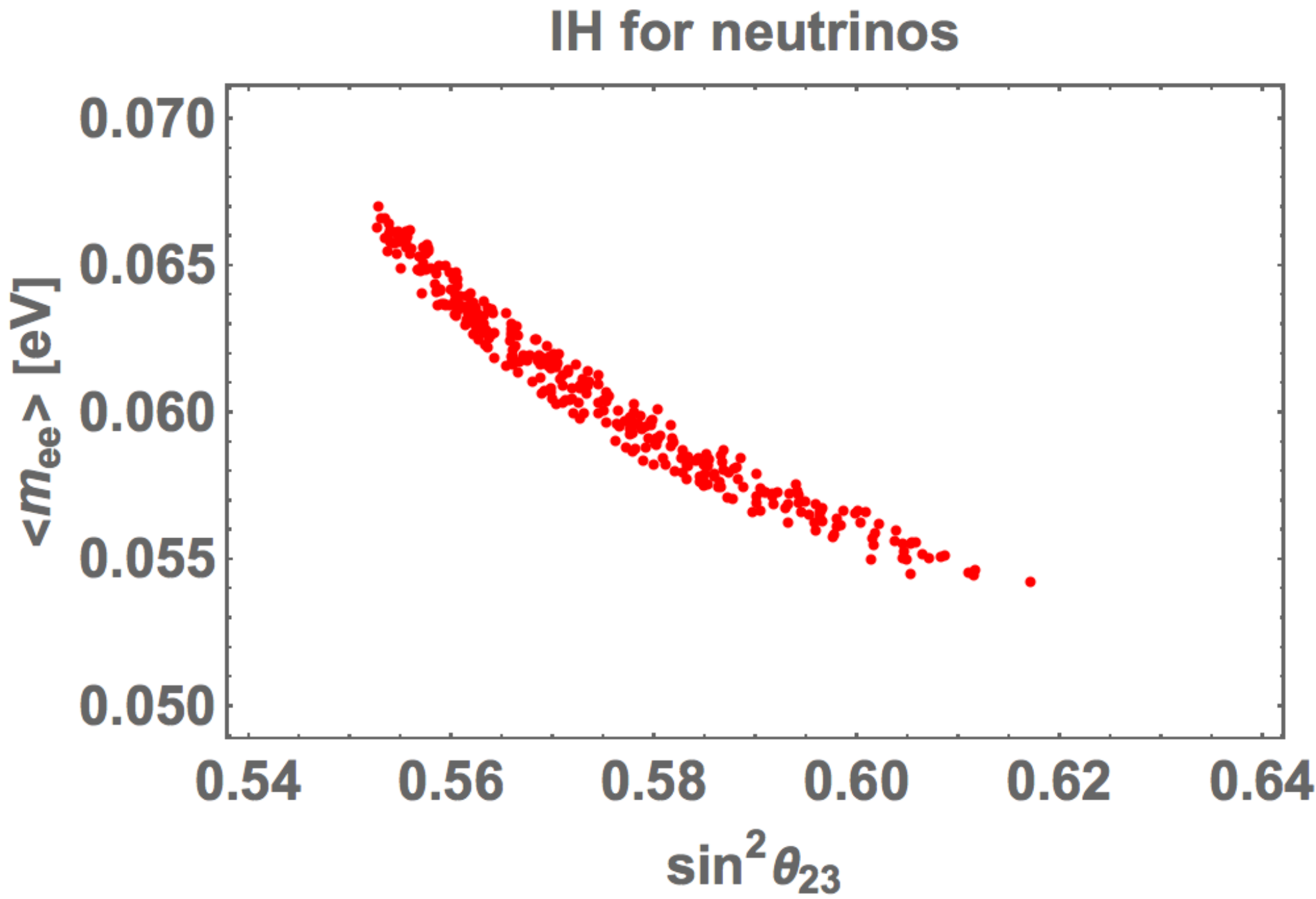}
		\caption{ The predicted 
			$\langle m_{ee}\rangle$ versus $\sin^2\theta_{23}$ for IH.}
	\end{minipage}
\end{figure}
%%%%%%%%%%%%%%%%%%%%%%%%%%%%%%%%%%%%%%%%%%%%%
%%%%%%%%%%%%%%%%%%%%%%%%%%%%%%%%%%%%%%%%%%%%%%
\subsection{Parameter samples of NH and IH}
 We show the numerical result of two samples for NH and IH, respectively.
%%%%%%%%%%%%%%%%%%%%%%%%%%%%%%%%%%%%%%%%%%%%%%%%%%%%%%%%%%%%%%%
In Table \ref{samplelepton},
parameters and outputs of our calculations
are presented for both NH and IH.
%%%%%%%%%%%%%%%%%%%%%%%%%%%%%%%%%%%%%%%%%%%%%%%%%%%
\begin{table}[H]
	\centering
	\begin{tabular}{|c|c|c|} \hline 
		\rule[14pt]{0pt}{0pt}	  &  NH&  IH\\ \hline 
		\rule[14pt]{0pt}{0pt}	
		$\tau$&   $ -0.0796 + 1.0065  \, i$  & $0.0103+ 1.0812 \, i$ \\ 
		\rule[14pt]{0pt}{0pt}
		$g^\nu_1$ &$ 0.124$ &-1.17\\
		\rule[14pt]{0pt}{0pt}
		$g^\nu_2$  &  $-0.802$ &6.79\\
		\rule[14pt]{0pt}{0pt}
		$\alpha_e/\gamma_e$ & $6.82\times 10^{-2}$  & $6.76\times 10^{-2}$\\
		\rule[14pt]{0pt}{0pt} 
		$\beta_e/\gamma_e$ &  $ 1.02\times 10^{-3}$  & $1.02\times 10^{-3}$\\
		\rule[14pt]{0pt}{0pt}
		$\sin^2\theta_{12}$ & $ 0.290$	& $ 0.291$\\
		\rule[14pt]{0pt}{0pt}
		$\sin^2\theta_{23}$ &  $ 0.564$	& $ 0.579$\\
		\rule[14pt]{0pt}{0pt}
		$\sin^2\theta_{13}$ &  $ 0.0225$	&  $ 0.0219$\\
		\rule[14pt]{0pt}{0pt}
		$\delta_{CP}^\ell$ &  $258^\circ$ 	&  $262^\circ$\\
		\rule[14pt]{0pt}{0pt}
		$[\alpha_{21},\,\alpha_{31}]$ &  $[330^\circ,\,338^\circ]$ 	&$[3.24^\circ,\,182^\circ]$\\	
		\rule[14pt]{0pt}{0pt}
		$\sum m_i$ &  $97.9$\,meV 	& $153$\,meV\\
		\rule[14pt]{0pt}{0pt}
		$\langle m_{ee} \rangle$ &  $19.2$\,meV 	&  $59.1$\,meV \\
		\rule[14pt]{0pt}{0pt}
		$\chi^2$ &  $1.98$ 	&  $4.12$ \\
		\hline
	\end{tabular}
	\caption{Numerical values of parameters and observables
		at the sample points of NH and IH.}
	\label{samplelepton}
\end{table}

 %%%%%%%%%%%%%%%%%%%%%%%%%%%%%%%%%%%%%%%%
 We also present the mixing matrices of   charged leptons $U_E$ and
 neutrinos $U_\nu$ for the  samples  of Table \ref{samplelepton}.
For NH, those are:
 \begin{align}
 \begin{aligned}
 U_E&\approx
 \begin{pmatrix}
0.983 & -0.020+0.158\, i & -0.011+0.092\, i\\
0.016+0.130\, i& 0.958 & -0.255+0.001\,  i\\
	0.016+0.129\, i& 0.239+0.001\, i &0.962 
 \end{pmatrix} \ , \\
 U_\nu&\approx
 \begin{pmatrix}
0.838 & -0.541+0.068\, i & -0.008+0.031\, i\\
0.450+0.076\, i& 0.688 & 0.564-0.0008\, i\\
-0.299-0.021\, i& -0.478-0.020\, i& 0.825
 \end{pmatrix} \ ,
 \end{aligned}
 \label{V-lepton}
 \end{align}
 which  are  given  in the diagonal base of the generator $S$
 in order to see  the hierarchical structure of flavor mixing \cite{Okada:2020ukr}.
  The PMNS mixing matrix is given as 
 $U_{\rm PMNS}=U_E^\dagger \, U_\nu$.
 The diagonal base of $S$ is obtained  by using the following unitary matrix:
  \begin{align}
 \begin{aligned}
 V_{S}\equiv \begin{pmatrix} 
 -\frac{1}{\sqrt{6}} &\ \frac{2}{\sqrt{6}} &  -\frac{1}{\sqrt{6}} \\
 \ \, \frac{1}{\sqrt{3}} &\frac{1}{\sqrt{3}} & \ \frac{1}{\sqrt{3}} \\
 -\frac{1}{\sqrt{2}} &0 & \ \frac{1}{\sqrt{2}}
 \end{pmatrix},
 \end{aligned}
 \label{Sdiagonal}
 \end{align}
 which  leads to
 $V_S\, S \, V_S^\dagger ={\rm diag }\, (1,\, -1,\, -1)$ \cite{Okada:2020ukr}.
 Then, the charged lepton and neutrino mass matrices are transformed as 
  $V_S M_f^\dagger M_f V_S^\dagger \, (f=E,\,\nu)$.
 %%%%%%%%%%%%%%%%%%%%%%%%%%%%%%%%%%

For IH, the mixing matrices are:
\begin{align}
\begin{aligned}
U_E&\approx
\begin{pmatrix}
0.983 & 0.155 +0.019 \,  i & 0.091+0.011\, i\\
	0.127+0.015 \, i& 0.956 & -0.264-0.001 \, i\\
-0.128+0.016 \, & 0.248-0.001 \, i & 0.960
\end{pmatrix} \ , \\
U_\nu&\approx
\begin{pmatrix}
0.840 & 0.0007+0.542 \, i & 0.032-0.001 \, i \\
-0.022+0.445 \,i & 0.691 & 0.570-0.002 \, i \\
-0.016-0.310 \, i & -0.478-0.023 \,i & 0.821
\end{pmatrix} \ ,
\end{aligned}
\label{V-leptonIH}
\end{align}
which  are also  given  in the diagonal base of the generator $S$.

 %%%%%%%%%%%%%%%%%%%%%%%%%%%%%%%%%%%%%%%%%%%%%%%%%%%%%%%%%%%
For both NH and IH, the  mixing matrix of  charged leptons $U_E$
 is hierarchical one,
 on the other hand, two large mixing angles of  1--2 and 2--3 flavors appear
 in the neutrino mixing matrix $U_\nu$.

%%%%%%%%%%%%%%%%%%%%%%%%%%%%%%%%%%%%%%%%%%%%%%%%%%%%%%%%%%%%%
%%%%%%%%%%%%%%%%%%%%%%%%%%%%%%%%%%%%%%%%%%%%%%%%%%%%%%%%%%%%%
%%%%%%%%%%%%%%%%%%%%%%%%  REG  %%%%%%%%%%%%%%%%%%%%%%%%%%%%%%
%%%%%%%%%%%%%%%%%%%%%%%%%%%%%%%%%%%%%%%%%%%%%%%%%%%%%%%%%%%%%

In our numerical calculations, we have not included  the RGE effects
 in the lepton mixing angles and neutrino mass ratio
 	$\Delta m_{\rm sol}^2/\Delta m_{\rm atm}^2$.
 We suppose that those corrections  are very small between 
 the electroweak  and GUT scales.
 % for NH of neutrino masses.
This assumption is  justified  well in the case of $\tan\beta\leq 5$
unless neutrino masses are almost degenerate \cite{Criado:2018thu}.

%%%%%%%%%%%%%%%%%%%%%%%%%%%%%%%%%%%%%%%%%%%%%%%%%%%%%%%%%%

\section{Summary and discussions}

The modular invariant $A_4$ model of lepton flavors  has been  studied combining with the generalized CP symmetry.
In our model, 
both CP and modular symmetries are broken spontaneously by VEV of the modulus $\tau$. The source of the CP violation is
 a non-trivial value of  ${\rm Re} [\tau]$ while parameters of neutrinos $g^\nu_1$ and $g^\nu_2$ are real.
 
 We have found allowed region of $\tau$ close to the  fixed point $\tau=i$,
  which is consistent with the observed lepton mixing angles and lepton masses
  for NH at  $2\,\sigma$   confidence level.
  The CP violating Dirac phase $\delta_{CP}$
  is predicted clearly in $[98^\circ,110^\circ]$ and  $[250^\circ,262^\circ]$
  at  $3\,\sigma$ confidence level. 
  The predicted $\sum m_i$ is in $[82,\,102]$\,meV with $3\,\sigma$ confidence level.
  
  %%%%%%%%%%%%%%%%%%%%%%%%%%%%%%%%%%%%%%%%%%%%%%%%%%%
  
   There is also  allowed region of $\tau$ close to the  fixed point $\tau=i$
   for IH  at  $5\,\sigma$   confidence level.
   The predicted  $\delta_{CP}$ is in 
   $[95^\circ,100^\circ]$ and  $[260^\circ,265^\circ]$ at $5\,\sigma$  confidence level.  The sum of neutrino masses is predicted in $\sum m_i=[134,\,180]$\,meV.

By using  the predicted Dirac phase and the Majorana phases,
we have obtained the effective mass 
$\langle m_{ee}\rangle$ for the $0\nu\beta\beta$ decay, which are
 in $[12.5,\,20.5]$\,meV for NH at $3\,\sigma$ confidence level
 and   in $[54,\,67]$\,meV for IH at  $5\,\sigma$ confidence level.
Since  KamLAND-Zen experiment \cite{KamLAND-Zen:2016pfg} 
 presented the upper bound on the effective Majorana mass as $\langle m_{ee}\rangle<61$--$165$\,meV 
 by using a variety of nuclear matrix element calculations,
 the prediction of  $[54,\,67]$\,meV for IH will be tested in the near future.
 Furthermore,
 the prediction of $\langle m_{ee}\rangle\simeq 20$\,meV for  NH will be also testable in the future experiments of the neutrinoless double beta decay.
 
 %%%%%%%%%%%%%%%%%%%%%%%%%%%%%%
  Since the CP symmetry is conserved at the boundary of the fundamental domain, one may expect the size of CP violation to be small 
  at the nearby fixed  point of $\tau=i$.
  In order to estimate of the size of CP violation, we can calculate the rephasing invariant CP violating measure of leptons, $J_{CP}$
  \cite{Jarlskog:1985ht,Krastev:1988yu}
 from mass matrices directly \cite{Shimizu:2019edl}.
  By using aproximate forms of lepton mass matrices at nearby fixed points
   in Ref.\cite{Okada:2020ukr}, we have obtained  the relation
   between  the magnitude of  $J_{CP}$ and  the deviation from $\tau=i$
   semi-quantitatively.
   In order to reproduce the almost maximal size $|J_{CP}|=0.03$,
  it is enough to take $\epsilon=\pm {\cal O} (0.05)$  where 
  $\epsilon$ is supposed to be real in the definition of  $\tau=i+\epsilon$.
  Since it is important to study   CP violation at nearby fixed points
  complehensively, we will present appropriate  forms in another paper.
 
 %%%%%%%%%%%%%%%%%%%%%%%%%%%%%%
 
 In our model,  the modulus $\tau$ dominates the CP violation.
 Therefore, the determination of  $\tau$ is the most important work.
Although we have constrained  $\tau$ by  observables of leptons phenomenologically,
one also should pay attention to the recent theoretical work of 
 the moduli stabilization from the viewpoint of modular  flavor symmetries
 \cite{1831045}. 
  The study of modulus $\tau$ is interesting to reveal the flavor theory in both theoretical and phenomenological aspects.
 
%%%%%%%%%%%%%%%%%%%%%%%%%%%%%%%%%%%%%%%%%%%%%%%%%%%%%%%%%%%%%
%%%%%%%%%%%%%%%%%%%%%%%%%%%%%%%%%%%%%%%%%%%%%%%%%%%%%%%%%%%%%
%%%%%%%%%%%%%%%%%%%%%%%%%%%%%%%%%%%%%%%%%%%%%%%%%%%%%%%%%%%%%
\section*{Acknowledgments}
This research was supported by an appointment to the JRG Program at the APCTP through the Science and Technology Promotion Fund and Lottery Fund of the Korean Government. This was also supported by
 the Korean Local Governments - Gyeongsangbuk-do Province and Pohang City (H.O.). H. O. is sincerely grateful for the KIAS member. 

%%%%%%%%%%%%%%%%%%%%%%%%%%%%%%%%%%%

%\newpage
\appendix
\section*{Appendix}

%%%%%%%%%%%%%%%%%%%%%%%%%%%%%%%%%%%%%%%%%%%%%%%%%%%%%%%%%%%
\section{Tensor product of  $A_4$ group}
%%%%%%%%%%%%%%%%%%%%%%%%%%%%%%%%%%%%%%%%%%%%%%%%%%%%%%%%%%%%%%%%%%%%%%%%
We take the generators of $A_4$ group for the triplet as follows:
\begin{align}
\begin{aligned}
S=\frac{1}{3}
\begin{pmatrix}
-1 & 2 & 2 \\
2 &-1 & 2 \\
2 & 2 &-1
\end{pmatrix},
\end{aligned}
\qquad 
\begin{aligned}
T=
\begin{pmatrix}
1 & 0& 0 \\
0 &\omega& 0 \\
0 & 0 & \omega^2
\end{pmatrix}, 
\end{aligned}
\end{align}
where $\omega=e^{i\frac{2}{3}\pi}$ for a triplet.
In this base,
the multiplication rule is
\begin{align}
\begin{pmatrix}
a_1\\
a_2\\
a_3
\end{pmatrix}_{\bf 3}
\otimes 
\begin{pmatrix}
b_1\\
b_2\\
b_3
\end{pmatrix}_{\bf 3}
&=\left (a_1b_1+a_2b_3+a_3b_2\right )_{\bf 1} 
\oplus \left (a_3b_3+a_1b_2+a_2b_1\right )_{{\bf 1}'} \nonumber \\
& \oplus \left (a_2b_2+a_1b_3+a_3b_1\right )_{{\bf 1}''} \nonumber \\
&\oplus \frac13
\begin{pmatrix}
2a_1b_1-a_2b_3-a_3b_2 \\
2a_3b_3-a_1b_2-a_2b_1 \\
2a_2b_2-a_1b_3-a_3b_1
\end{pmatrix}_{{\bf 3}}
\oplus \frac12
\begin{pmatrix}
a_2b_3-a_3b_2 \\
a_1b_2-a_2b_1 \\
a_3b_1-a_1b_3
\end{pmatrix}_{{\bf 3}\  } \ , \nonumber \\
\nonumber \\
{\bf 1} \otimes {\bf 1} = {\bf 1} \ , \qquad &
{\bf 1'} \otimes {\bf 1'} = {\bf 1''} \ , \qquad
{\bf 1''} \otimes {\bf 1''} = {\bf 1'} \ , \qquad
{\bf 1'} \otimes {\bf 1''} = {\bf 1} \  ,
\end{align}
where
\begin{align}
T({\bf 1')}=\omega\,,\qquad T({\bf 1''})=\omega^2. 
\end{align}
More details are shown in the review~\cite{Ishimori:2010au,Ishimori:2012zz}.

%%%%%%%%%%%%%%%%%%%%%%%%%%%%%%%%%%%%%%%%%%%%%%%%%%%%%%%%%%%
\section{Modular forms in $A_4$ symmetry}
%%%%%%%%%%%%%%%%%%%%%%%%%%%%%%%%%%%%%%%%
For $\Gamma_3\simeq A_4$, the dimension of the linear space 
${\cal M}_k(\Gamma{(3)})$ 
of modular forms of weight $k$ is $k+1$ \cite{Gunning:1962,Schoeneberg:1974,Koblitz:1984}, i.e., there are three linearly 
independent modular forms of the lowest non-trivial weight $2$.
These forms have been explicitly obtained \cite{Feruglio:2017spp} in terms of
the Dedekind eta-function $\eta(\tau)$: 
%which is written by 
%%%%%%%%%%%%%%%%%%%%%%%%%%%%%
\begin{equation}
\eta(\tau) = q^{1/24} \prod_{n =1}^\infty (1-q^n)~, 
\quad\qquad  q= \exp \ (i 2 \pi  \tau )~,
\label{etafunc}
\end{equation}
%%%%%%%%%%%%%%%%%%%%%%%%%%
%
where $\eta(\tau)$ is a  so called  modular form of weight~$1/2$. 
% and $\eta(\tau)$ is a modular form of weight~$1/2$.
In what follows we will use the following base of the 
$A_4$ generators  $S$ and $T$ in the triplet representation:
%%%%%%%%%%%%%%%%%%%%%%%%%%%
\begin{align}
\begin{aligned}
S=\frac{1}{3}
\begin{pmatrix}
-1 & 2 & 2 \\
2 &-1 & 2 \\
2 & 2 &-1
\end{pmatrix},
\end{aligned}
\qquad \qquad
\begin{aligned}
T=
\begin{pmatrix}
1 & 0& 0 \\
0 &\omega& 0 \\
0 & 0 & \omega^2
\end{pmatrix}, 
\end{aligned}
\label{STbase0}
\end{align}
%%%%%%%%%%%%%%%%%%%%%%%%%%%%%%%%%
%
where $\omega=\exp (i\frac{2}{3}\pi)$ .
The  modular forms of weight 2 $(k=2)$ transforming
as a triplet of $A_4$, ${\bf Y^{\rm (2)}_3}(\tau)=(Y_1(\tau)\, Y_2(\tau), Y_3(\tau))^T$, can be written in terms of 
$\eta(\tau)$ and its derivative \cite{Feruglio:2017spp}:
%%%%%%%%%%%%%%%%%%%%%%%
\begin{eqnarray} 
\label{eq:Y-A40}
Y_1(\tau) &=& \frac{i}{2\pi}\left( \frac{\eta'(\tau/3)}{\eta(\tau/3)}  +\frac{\eta'((\tau +1)/3)}{\eta((\tau+1)/3)}  
+\frac{\eta'((\tau +2)/3)}{\eta((\tau+2)/3)} - \frac{27\eta'(3\tau)}{\eta(3\tau)}  \right), \nonumber \\
Y_2(\tau) &=& \frac{-i}{\pi}\left( \frac{\eta'(\tau/3)}{\eta(\tau/3)}  +\omega^2\frac{\eta'((\tau +1)/3)}{\eta((\tau+1)/3)}  
+\omega \frac{\eta'((\tau +2)/3)}{\eta((\tau+2)/3)}  \right) , \label{Yi} \\ 
Y_3(\tau) &=& \frac{-i}{\pi}\left( \frac{\eta'(\tau/3)}{\eta(\tau/3)}  +\omega\frac{\eta'((\tau +1)/3)}{\eta((\tau+1)/3)}  
+\omega^2 \frac{\eta'((\tau +2)/3)}{\eta((\tau+2)/3)}  \right)\,.
\nonumber
\end{eqnarray}
%%%%%%%%%%%%%%%%%%%%%
%
% where 
The overall coefficient in Eq.\,(\ref{Yi}) is 
one possible choice.
% and cannot be determined essentially.
It cannot be uniquely determined.
The triplet modular forms of weight $2$
% are also  expressed in the $q$ expansions
have the following  $q$-expansions:
%%%%%%%%%%%%%%%%%%%%%%%%%%
\begin{align}
{\bf Y^{\rm (2)}_3}(\tau)
=\begin{pmatrix}Y_1(\tau)\\Y_2(\tau)\\Y_3(\tau)\end{pmatrix}=
\begin{pmatrix}
1+12q+36q^2+12q^3+\dots \\
-6q^{1/3}(1+7q+8q^2+\dots) \\
-18q^{2/3}(1+2q+5q^2+\dots)\end{pmatrix}.
\label{Y(2)}
\end{align}
%%%%%%%%%%%%%%%%%%%%%%%
%
% where $Y_i^{(2)}(\tau)$ 
They satisfy also the constraint \cite{Feruglio:2017spp}:
%%%%%%%%%%%%%%%%%%%%%%%%
\begin{align}
Y_2(\tau)^2+2Y_1(\tau) Y_3(\tau)=0~.
\label{condition}
\end{align}
%%%%%%%%%%%%%%%%%%%%%
%%%%%%%%%%%%%%%%%%%%%
%%%%%%%%%%%%%%%%%%%%

The  modular forms of the  higher weight, $k$, can be obtained
by the $A_4$ tensor products of  the modular forms  with weight 2,
${\bf Y^{\rm (2)}_3}(\tau)$, 
as given in Appendix A.
For weight 4, that is $k=4$, there are  five modular forms
by the tensor product of  $\bf 3\otimes 3$ as:
\begin{align}
&\begin{aligned}
{\bf Y^{\rm (4)}_1}(\tau)=Y_1(\tau)^2+2 Y_2(\tau) Y_3(\tau) \, , \qquad\quad\ \
{\bf Y^{\rm (4)}_{1'}}(\tau)=Y_3(\tau)^2+2 Y_1(\tau) Y_2(\tau) \, , 
\end{aligned}\nonumber \\
\nonumber \\
&\begin{aligned} 
{\bf Y^{\rm (4)}_{1''}}(\tau)=Y_2(\tau)^2+2 Y_1(\tau) Y_3(\tau)=0 \, , \qquad
{\bf Y^{\rm (4)}_{3}}(\tau)=
\begin{pmatrix}
Y_1^{(4)}(\tau)  \\
Y_2^{(4)}(\tau) \\
Y_3^{(4)}(\tau)
\end{pmatrix}
=
\begin{pmatrix}
Y_1(\tau)^2-Y_2(\tau) Y_3(\tau)  \\
Y_3(\tau)^2 -Y_1(\tau) Y_2(\tau) \\
Y_2(\tau)^2-Y_1(\tau) Y_3(\tau)
\end{pmatrix}\, , 
\end{aligned}
\label{weight4}
\end{align}
where ${\bf Y^{\rm (4)}_{1''}}(\tau)$ vanishes due to the constraint of
Eq.\,(\ref{condition}).

%%%%%%%%%%%%%%%%%%%%%%%%%%%%%
%For weight $6$, there are  seven modular forms
%by the tensor products of  $A_4$ as:
%\begin{align}
%&\begin{aligned}
%{\bf Y^{(6)}_1}=Y_1^3+ Y_2^3+Y_3^3 -3Y_1 Y_2 Y_3  \ , 
%\end{aligned} \nonumber \\
%\nonumber \\
%&\begin{aligned} {\bf Y^{(6)}_3}\equiv 
%\begin{pmatrix}
%Y_1^{(6)}  \\
%Y_2^{(6)} \\
%Y_3^{(6)}
%\end{pmatrix}
%=
%\begin{pmatrix}
%Y_1^3+2 Y_1 Y_2 Y_3   \\
%Y_1^2 Y_2+2 Y_2^2 Y_3 \\
%Y_1^2Y_3+2Y_3^2Y_2
%\end{pmatrix}\ , \qquad
%\end{aligned}
%\begin{aligned} {\bf Y^{(6)}_{3'}}\equiv
%\begin{pmatrix}
%Y_1^{'(6)}  \\
%Y_2^{'(6)} \\
%Y_3^{'(6)}
%\end{pmatrix}
%=
%\begin{pmatrix}
%Y_3^3+2 Y_1 Y_2 Y_3   \\
%Y_3^2 Y_1+2 Y_1^2 Y_2 \\
%Y_3^2Y_2+2Y_2^2Y_1
%\end{pmatrix}\ . 
%\end{aligned}
%\label{weight6}
%\end{align}
%By using these modular forms of weights $2, 4$ and $6$,
%we discuss  lepton mass matrices.

%%%%%%%%%%%%%%%%%%%%%%%%%%%%%%%%%%%%%%%%%%%%%%%
\section{Determination of $\alpha_e/\gamma_e$ and $\beta_e/\gamma_e$}

The coefficients $\alpha_e$, $\beta_e$, and $\gamma_e$ in Eq.(\ref{ME(2)})
are taken to be real positive without loss of generality.
We show these parameters are described in terms of the modular parameter $\tau$ and the charged lepton masses.
We rewrite the mass matrix of Eq.\,(\ref{ME(2)}) as  
 \begin{align}
\begin{aligned}
M_E=v_d \gamma_e
\begin{pmatrix}
\hat\alpha_e & 0 & 0 \\
0 & \hat\beta_e & 0\\
0 & 0 &1
\end{pmatrix} 
\begin{pmatrix}
Y_1(\tau) & Y_3(\tau)& Y_2(\tau) \\
Y_2(\tau) & Y_1(\tau)&  Y_3(\tau) \\
Y_3(\tau)&  Y_2(\tau)&  Y_1(\tau)
\end{pmatrix}
\, ,     
\end{aligned}
\label{matrixB}
\end{align}
where $\hat{\alpha}_e\equiv\alpha_e/\gamma_e$ and  $\hat{\beta}_e\equiv\beta_e/\gamma_e$.
Denoting charged lepton masses $m_1=m_e$, $m_2=m_\mu$ and  $m_3=m_\tau$,
 we have three equations as:
\begin{align}
%\sum_{i=e}^\tau m_i^2
{\sum_{i=1}^3 m_{i}^2}
={\rm Tr}[M_E^\dag M_E]&=v_d^2\gamma_e^2 \ (1+\hat\alpha_e^2+\hat\beta_e^2)\ C^e_{1}\,,\label{eq:sum} \\
%%%%%%%%%%%%%%%%%%%
%\prod_{i=e}^\tau m_i^2
{\prod_{i=1}^3 m_{i}^2}
={\rm Det}[M_E^{\dag} M_E]&=
v_d^6\gamma_e^6\  \hat\alpha^2_e\hat\beta^2_e \ C^e_{2}\,, \label{eq:prod}\\
%%%%%%%%%%%%%%%%%%%%%%%%%%%%%%%%%%
\chi= \frac{{\rm Tr}[M_E^{\dag} M_E]^2-{\rm Tr}[(M_E^{\dag} M_E)^2]}{2}
&=v_d^4\gamma^4_e\  (\hat\alpha_e^2+\hat\alpha^2_e\hat\beta_e^2+\hat\beta_e^2)~C^e_{3}\,, \label{eq:chi}
\end{align}
where
{$\chi\equiv m_{1}^2m_{2}^2+m_{2}^2m_{3}^2+m_{3}^2m_{1}^2$}.
The coefficients $C^e_{1}$, $C^e_{2}$ and $C^e_{3}$ depend only on $Y_i(\tau)$'s,
 where
$Y_i(\tau)$'s  are determined if the value of modulus $\tau$ is fixed.
Those are given explicitly as follows:
\begin{align}
\begin{aligned}
C^e_{1}&=| Y_1(\tau)|^2+| Y_2(\tau)|^2+| Y_3(\tau)|^2\,, \nonumber \\
C^e_{2}&= | Y_1(\tau)^3 + Y_2(\tau)^3 +  Y_3(\tau)^3 
-3  Y_1(\tau)  Y_2(\tau)  Y_3(\tau)|^2\,,
\nonumber \\
C^e_{3}&= | Y_1(\tau)|^4+| Y_2(\tau)|^4+|Y_3(\tau)|^4
+| Y_1(\tau) Y_2(\tau)|^2+| Y_2(\tau) Y_3(\tau)|^2+
| Y_1(\tau) Y_3(\tau)|^2 \nonumber\\
&-2{\rm Re}\left [ Y_1^*(\tau)  Y_2^*(\tau) Y_3^2(\tau) 
+ Y_1^2(\tau)  Y_2^*(\tau) Y_3^*(\tau)+  Y_1^*(\tau) Y_2^2(\tau) Y_3^*(\tau) \right ] \,.
\end{aligned}
\end{align}

Then, we obtain two equations
which describe $\hat\alpha_e$ and $\hat\beta_e$ in terms of   masses and $\tau$:
\begin{align}
\begin{aligned}
\frac{(1+s)(s+t)}t&=\frac{(\sum m_i^2/C^e_1)(\chi/C^e_3)}{\prod m_i^2/C^e_2}~,\quad\qquad
\frac{(1+s)^2}{s+t}&=\frac{(\sum m_i^2/C^e_1)^2}{\chi/C^e_3}~,
\end{aligned}
\end{align}
where we redefine the parameters $\hat\alpha_e^2+\hat\beta_e^2=s$ and $\hat\alpha_e^2\hat\beta_e^2=t$.
After fixing charged lepton masses and $\tau$, we obtain $s$ and $t$
 numerically.
They are related as follows:
\begin{align}
\hat\alpha_e^2=\frac{s\pm\sqrt{s^2-4t}}2\,,\quad\quad
\hat\beta_e^2=\frac{s\mp\sqrt{s^2-4t}}2\,.
\label{alphabeta}
\end{align}

%%%%%%%%%%%%%%%%%%%%%%%%%%%%%%%%%%%%%%%%%%%%%%%%%%%%%%%%%
%%%%%%%%%%%%%%%%%%%%%%%%%%%%%%%%%%%%%%%%%%%%%%%%%%%%%%%%%
\section{Majorana and Dirac phases and $\langle m_{ee}\rangle $
in  $0\nu\beta\beta$ decay }

Supposing neutrinos to be Majorana particles, 
the PMNS matrix $U_{\text{PMNS}}$~\cite{Maki:1962mu,Pontecorvo:1967fh} 
is parametrized in terms of the three mixing angles $\theta _{ij}$ $(i,j=1,2,3;~i<j)$,
one CP violating Dirac phase $\delta _\text{CP}$ and two Majorana phases 
$\alpha_{21}$, $\alpha_{31}$  as follows:
\begin{align}
U_\text{PMNS} =
\begin{pmatrix}
c_{12} c_{13} & s_{12} c_{13} & s_{13}e^{-i\delta_\text{CP}} \\
-s_{12} c_{23} - c_{12} s_{23} s_{13}e^{i\delta_\text{CP}} &
c_{12} c_{23} - s_{12} s_{23} s_{13}e^{i\delta_\text{CP}} & s_{23} c_{13} \\
s_{12} s_{23} - c_{12} c_{23} s_{13}e^{i\delta_\text{CP}} &
-c_{12} s_{23} - s_{12} c_{23} s_{13}e^{i\delta_\text{CP}} & c_{23} c_{13}
\end{pmatrix}
%\times
\begin{pmatrix}
1&0 &0 \\
0 & e^{i\frac{\alpha_{21}}{2}} & 0 \\
0 & 0 & e^{i\frac{\alpha_{31}}{2}}
\end{pmatrix},
\label{UPMNS}
\end{align}
where $c_{ij}$ and $s_{ij}$ denote $\cos\theta_{ij}$ and $\sin\theta_{ij}$, respectively.

The rephasing invariant CP violating measure of leptons \cite{Jarlskog:1985ht,Krastev:1988yu}
is defined by the PMNS matrix elements $U_{\alpha i}$. 
It is written in terms of the mixing angles and the CP violating phase as:
\begin{equation}
J_{CP}=\text{Im}\left [U_{e1}U_{\mu 2}U_{e2}^\ast U_{\mu 1}^\ast \right ]
=s_{23}c_{23}s_{12}c_{12}s_{13}c_{13}^2\sin \delta_\text{CP}\,,
\label{Jcp}
\end{equation}
where $U_{\alpha i}$ denotes the each component of the PMNS matrix.

There are also other invariants $I_1$ and $I_2$ associated with Majorana phases
%\cite{Bilenky:2001rz}-\cite{Girardi:2016zwz},
\begin{equation}
I_1=\text{Im}\left [U_{e1}^\ast U_{e2} \right ]
=c_{12}s_{12}c_{13}^2\sin \left (\frac{\alpha_{21}}{2}\right )\,, \quad
I_2=\text{Im}\left [U_{e1}^\ast U_{e3} \right ]
=c_{12}s_{13}c_{13}\sin \left (\frac{\alpha_{31}}{2}-\delta_\text{CP}\right )\,.
\label{Jcp}
\end{equation}
We can calculate $\delta_\text{CP}$, $\alpha_{21}$ and $\alpha_{31}$ with these relations by taking account of 
\begin{eqnarray}
&&\cos\delta_{CP}=\frac{|U_{\tau 1}|^2-
	s_{12}^2 s_{23}^2 -c_{12}^2c_{23}^2s_{13}^2}
{2 c_{12}s_{12}c_{23}s_{23}s_{13}}\,, \nonumber \\
&&\text{Re}\left [U_{e1}^\ast U_{e2} \right ]
=c_{12}s_{12}c_{13}^2\cos \left (\frac{\alpha_{21}}{2}\right )\,, \qquad
\text{Re}\left [U_{e1}^\ast U_{e3} \right ]
=c_{12}s_{13}c_{13}\cos\left(\frac{\alpha_{31}}{2}-\delta_\text{CP}\right )\,.
\end{eqnarray}
In terms of this parametrization, the effective mass for the $0\nu\beta\beta$ decay is given as follows:
\begin{align}
\langle m_{ee}	\rangle=\left| m_1 c_{12}^2 c_{13}^2+ m_2s_{12}^2 c_{13}^2 e^{i\alpha_{21}}+
 m_3 s_{13}^2 e^{i(\alpha_{31}-2\delta_{CP})}\right|  \, .
\end{align}

%%%%%%%%%%%%%%%%%%%%%%%%%%%%%%%%%%%%%%%%%%%%%%%%%%%%%%%%%%
%%%%%%%%%%%%%%%%%%%%%%%%%%%%%%%%%%%%%%%%%%%%%%%%%%%%%%%%%%
%%%%%%%%%%%%%%%%%%%%%   References   %%%%%%%%%%%%%%%%%%%%% %%%%%%%%%%%%%%%%%%%%%%%%%%%%%%%%%%%%%%%%%%%%%%%%%%%%%%%%%%%
%\newpage

\end{document}